\documentclass[12pt,a4paper]{article}
\usepackage[left=1in,top=1in,right=1in,nohead]{geometry}
\usepackage{enumerate}
\usepackage{graphicx}
\usepackage{subfig}
\usepackage{amsmath}
\usepackage{amssymb}
\usepackage{mathrsfs}
\usepackage{multirow}
\usepackage{bm}
\usepackage{color}
\usepackage{jheppub}
\usepackage{verbatim}
\usepackage{placeins}

\def\bea#1\eea{\begin{align}#1\end{align}}

\newcommand{\eps}{\epsilon}

\newcommand{\zt}{\tilde{z}}

\newcommand{\qt}{\tilde{q}}
\newcommand{\Q}{\mathcal{Q}}
\newcommand{\E}{\mathcal{E}}

\newcommand{\arctanh}{\mathrm{arctanh}}

\newcommand{\Tr}{\mathrm{Tr}}

\renewcommand{\Im}{\mathrm{Im}}

\newcommand{\cai}{ {\cal A}_1 }
\newcommand{\caii}{ {\cal A}_2 }

\begin{document}

\title{Entanglement and Correlations near Extremality: CFTs dual to Reissner-Nordstr\"om AdS${}_5$}

\author[a]{Tom\'as Andrade}
\author[b,c]{Sebastian Fischetti}
\author[b,c]{Donald Marolf}
\author[a]{Simon F. Ross}
\author[d]{Moshe Rozali}

\affiliation[a]{Centre for Particle Theory, Department of Mathematical Sciences \\ Durham University, South Road, Durham DH1 3LE, U.K.}
\affiliation[b]{Department of Physics \\ University of California,
Santa Barbara, Santa Barbara, CA 93106, USA}
\affiliation[c]{Department of Applied Mathematics and Theoretical Physics \\ University of Cambridge, CB3 0WA, UK}
\affiliation[d]{Department of Physics and Astronomy \\
University of British Columbia, Vancouver, BC V6T 1Z1, Canada}

\emailAdd{tomas.andrade@durham.ac.uk}
\emailAdd{sfischet@physics.ucsb.edu}
\emailAdd{marolf@physics.ucsb.edu}
\emailAdd{s.f.ross@durham.ac.uk}
\emailAdd{rozali@phas.ubc.ca}


\arxivnumber{1312.????}

\abstract{
We use the AdS/CFT correspondence to study models of entanglement and correlations between two $d=4$ CFTs in thermofield double states at finite chemical potential.  Our bulk spacetimes are planar Reissner-Nordstr\"om AdS black holes.  We compute both thermo-mutual information and the two-point correlators of large-dimension scalar operators, focussing on the small-temperature behavior -- an infrared limit with behavior similar to that seen at large times.   The interesting feature of this model is of course that the entropy density remains finite as $T \rightarrow 0$ while the bulk geometry develops an infinite throat.  This leads to a logarithmic divergence in the scale required for non-zero mutual information between equal-sized strips in the two CFTs, though the mutual information between one entire CFT and a finite-sized strip in the other can remain non-zero even at $T=0$.  Furthermore, despite the infinite throat, there can be extremally charged operators for which the two-point correlations remain finite as $T \rightarrow 0$.  This suggests an interestingly mixed picture in which some aspects of the entanglement remain localized on scales set by the chemical potential, while others shift  to larger and larger scales.  We also comment on implications for the localized-quasiparticle picture of entanglement.
}

\maketitle

%================================================================
\section{Introduction}
\label{sec:intro}
%================================================================

\begin{figure}
\centering
\includegraphics[page=1]{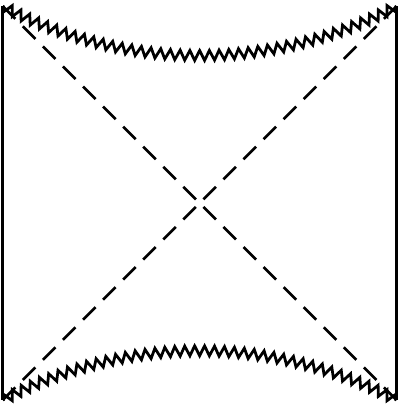}
\caption{A conformal diagram of the maximally extended planar AdS-Schwarzschild black hole.  This geometry is the bulk dual to the TFD state of two disconnected CFTs living on the two boundaries of the spacetime.}
\label{fig:Schwarzschild}
\end{figure}

The entanglement properties of both ground states and thermal states are interesting topics of study in quantum field theory; see e.g. \cite{Holzhey:1994we,Calabrese:2004eu}.  Here we consider the corresponding structure of so-called thermofield double (TFD) states, which are the natural pure states defined on an (essentially) identical pair of field theories that reduce to thermal density matrices on either theory alone.  This entanglement is of particular interest in the holographic context~\cite{Maldacena:1997re,Gubser:1998bc,Witten:1998qj}  due the existence in the dual bulk solution \cite{Maldacena:2001kr}  of a wormhole, or Einstein-Rosen-like bridge, between two asymptotic regions -- see figure \ref{fig:Schwarzschild} -- and the conjectured generalizations of \cite{VanRaamsdonk:2009ar,VanRaamsdonk:2010pw,Maldacena:2013xja}.    Holography also provides useful tools for such studies, ranging from the minimal area entanglement prescription of Ryu-Takayangi \cite{Ryu:2006bv} (recently justified in \cite{Lewkowycz:2013nqa}) to the particle-worldline approximation (a.k.a. the geodesic approximation) of bulk correlators.  TFDs also provides a simple laboratory in which to explore more general issues of entanglement in quantum field theory and holography.

We therefore focus on the holographic setting below.  TFD entanglement along these lines was explored in great detail for $d=2$ holographic CFTs in \cite{Morrison:2012iz} by investigating the dual BTZ black hole, and also earlier in more general contexts through studies of local two-point functions with one operator near each boundary of various two-sided black holes \cite{Maldacena:2001kr,Kraus:2002iv,Fidkowski:2003nf,Brecher:2004gn,Festuccia:2005pi,Hartman:2013qma}.  Our interest lies in adding charge via an appropriate chemical potential $\mu$ and exploring the behavior at very small temperatures $T$.

Without the chemical potential, taking $T \rightarrow 0$ simply drives each theory into its ground state and removes all correlations.  But nonzero $\mu$ provides an opportunity to maintain finite entanglement even at very small $T$.  The classic gravitational example of such behavior is of course the Reissner-Nordstr\"om black hole near extremality. For simplicity we therefore focus on TFD states which are holographically dual to planar Reissner-Nordstr\"om AdS (RNAdS).  To be concrete, we work with $d=4$ CFTs dual to $5$-dimensional bulks.

The interesting feature of such models is that as $T \rightarrow 0$ the bulk geometry develops a throat of finite cross-section but infinite depth.  The infinite depth leads many natural probes of entanglement to vanish at $T=0$.  For example, this is the case for two-point functions (with one argument in each CFT) of large-dimension neutral single-trace operators; such correlators decay exponentially with spacelike separation in the bulk. It is also the case for the mutual information between finite-sized regions of our two CFTs as computed via \cite{Ryu:2006bv}, as the diverging distance through the extreme throat means that at low $T$ the dominant contribution to the von Neumann entropy of any finite region is given by surfaces lying entirely on one side of the black hole.

Nevertheless, the total density of entanglement remains finite.  We take some first steps toward probing its structure below, showing in particular that i) the mutual information between one entire CFT and a finite-size strip in the other CFT need not vanish at small $T$ and ii) as suggested in \cite{Marolf:2013dba}, there can be what one may call extremally-charged operators whose two-point functions (with one argument in each CFT) remain finite in the $T \to 0$ limit. The existence of the above extremally-charged operators indicates that the system lies at the threshold of an instability of the extreme RNAdS spacetime associated with Schwinger pair creation\footnote{\label{instab} Since we consider bosons, this may also be called either a super-radiant instability or an instability to forming a super-conducting phase as in \cite{Gubser:2008px,Gubser:2008pf,Hartnoll:2008vx,Hartnoll:2008kx,Denef:2009tp}.  The fermionic analogue would be unstable to forming a Fermi surface as in \cite{Lee:2008xf,Liu:2009dm,Cubrovic:2009ye,Faulkner:2009wj}.}~\cite{Schwinger:1951nm,Gibbons:1975kk}.

After a brief review of TFDs and the RNAdS geometry in section \ref{sec:background}, we proceed to study the above mutual information in section \ref{sec:mutualinfo}.    Section \ref{sec:corr} then examines the two-point functions of {\it charged} operators with one argument in each CFT.     Some final discussion is given in section \ref{sec:discussion}, which in particular connects phenomena described here at small $T$ with similar infrared (IR) effects seen in \cite{Hartman:2013qma,Liu:2013iza,Liu:2013qca} at large times.

Before beginning, we mention the well-known fact that RNAdS has many potential instabilities that can switch on at low temperature (see e.g. \cite{Gubser:2008px,Hartnoll:2008vx,Hartnoll:2008kx,Lee:2008xf,Liu:2009dm,Cubrovic:2009ye,Faulkner:2009wj,Hartnoll:2009ns}), and one certainly does not expect the extreme limit of RNAdS\footnote{Or, in fact, any black hole whether extreme or otherwise; see e.g. \cite{Festuccia:2005pi} for a modern statement of this issue.} to give an exact description of any microscopic theory with a finite density of states \cite{Mathur:2005zp,Sen:2011cn,Jensen:2011su}.  But at any given $T/\mu$, even very close to extremality, models may well exist in which RNAdS remains an accurate description.   Furthermore, we expect our results to be typical of those obtained near extreme limits.  In particular, at least at first pass one would expect rotating extreme global AdS black holes to behave similarly.  In this context one can find black holes that saturate a BPS bound (extreme BTZ \cite{Banados:1992wn,Banados:1992gq} for AdS${}_3$ and the solutions of  \cite{Kostelecky:1995ei}  and \cite{Gutowski:2004ez} for AdS${}_4$ and AdS${}_5$), so they are free of the above supergravity instabilities.

%================================================================
\section{Thermofield Doubles in Bulk and CFT}
\label{sec:background}
%================================================================

We begin with a brief review of charged thermofield double states, both in the CFT and in the bulk.  In the latter context they become two-sided planar Reissner-Nordstr\"om AdS black holes.

\subsection{The Charged Thermofield Double in the CFT}
\label{sec:TFDCFT}

Consider two quantum systems with isomorphic Hilbert spaces ${\cal H}_1 = {\cal H}_2 = {\cal H}$ and identical Hamiltonians $H_1=H_2=H$, which for simplicity we take to be invariant under a time-reversal operation\footnote{\label{Trev} This will be the case for our system.  More generally, when time-reversal is not a symmetry, one takes $H_1$ and $H_2$ to be related by time-reversal with corresponding changes in \eqref{eq:TFD}.} ${\cal T}$.  We will be interested in considering this theory in the TFD state
\begin{equation}
\label{eq:TFD}
\left| \psi \right\rangle = \frac{1}{\sqrt{Z}} \sum_{i} e^{-\beta E_i/2} \left| E_i\right\rangle_1 \otimes \left| E_i\right\rangle_2 .
\end{equation}
Up to the insertion of possible phases, when the spectrum is non-degenerate this is the unique state on the tensor product ${\cal H}_1 \otimes {\cal H}_2$ which restricts to the thermal state
$\rho = \frac{1}{Z} \, e^{-\beta H}$ on each factor, where $Z = \Tr_{\cal H} e^{-\beta H}$ is the usual partition function.  The state \eqref{eq:TFD} may be constructed by cutting open the thermal path integral with inverse temperature $\beta$, where the cut is made along a surface invariant under time-reversal.  Equivalently, it may be evaluated by performing the Euclidean path integral in which Euclidean time $t_E$ runs over an interval $I$ of length $\beta/2$. Even in the presence of degeneracies, this Euclidean recipe continues to be well-defined, and implies that the terms in \eqref{eq:TFD} take the form $|E \rangle \otimes {\cal T} |E \rangle$. Writing the TFD in this form makes clear that constructing \eqref{eq:TFD} involves choosing a special time $t=0$ invariant under ${\cal T}$ and furthermore that, once this time has been chosen, the anti-linear nature of ${\cal T}$ makes the properly defined \eqref{eq:TFD} independent of changes of phase in the basis states $|E \rangle$. We also see that in relativistic theories CPT-invariance implies that  the two factors in \eqref{eq:TFD} should be taken to have opposite charge .

The generalization to the grand canonical ensemble is straightforward: we simply introduce a chemical potential into the Boltzmann weights and define the TFD state to be
\begin{equation}
\label{eq:TFDmu}
\left| \psi \right\rangle = \frac{1}{\sqrt{\mathcal{Z}}} \sum_{i} e^{-\beta (E_i+\mu Q_i)/2} \left| E_i,  Q_i\right\rangle_1 \otimes \left| E_i, -Q_i\right\rangle_2,
\end{equation}
where the~$Q_i$ are eigenvalues of the conserved $U(1)$ charge conjugate to~$\mu$ and~$\mathcal{Z}$ is now the grand partition function.  Again, any ambiguities due to degeneracies are resolved by taking the two states in each term to be CPT conjugates.  The state \eqref{eq:TFDmu} arises from a Euclidean path integral as above if we couple the charge $Q$ to a background~$U(1)$ gauge field $A = - i \mu \, dt_E$ with the sign in \eqref{eq:TFDmu} requiring us to take system 1 to be associated with the minimum value of $t_E$ in $I$ and system 2 to be associated with the maximum value.  Note that the gauge field is imaginary, and that the result is just the TFD state defined by the non-time-reversal invariant deformed Hamiltonians $\tilde H_1 = H + \mu Q$, $\tilde H_2 = H - \mu Q$; see footnote \ref{Trev}.

The TFD state \eqref{eq:TFDmu} has a well-behaved zero-temperature limit $\beta \to \infty$ only if $\tilde E_1 = E + \mu Q$ is bounded below, or equivalently (by applying ${\cal T}$) if $\tilde E_2 = E - \mu Q$ is bounded below.  In the zero-temperature limit, the sum in~\eqref{eq:TFDmu} restricts to those terms that minimize $\tilde E_1, \tilde E_2$. For a general theory one may expect a unique state of minimal $\tilde E_1,\tilde E_2$.  But symmetry can force an exact degeneracy or, alternatively, we may consider a theory with many degrees of freedom (e.g., large $N$) and an associated approximate degeneracy when $\beta$ is large but finite.  It is this latter option that one expects to apply to the RN-AdS black holes studied below (see e.g. comments in \cite{Jensen:2011su}).  In either case, up to an irrelevant overall phase the state becomes effectively independent of time evolutions generated by $\tilde H_1, \tilde H_2$.

Since any remaining entanglement is associated with excitations of vanishingly small energy above the ground state (in the sense of $\tilde H_1, \tilde H_2$), one might expect any spatial scale characterizing our TFD entanglement to diverge as $T \rightarrow 0$.  But this will not quite be the case.  Indeed, since we consider RNAdS${}_5$, our bulk dual will have an AdS${}_2 \times {\mathbb R}^3$ infrared fixed point describing the near-horizon region.  Such spacetimes exhibit local criticality, characterized by the limit $z \rightarrow \infty$ of dynamical scaling symmetry
$(t, x) \rightarrow (\lambda^z t, \lambda x)$ which for finite $z$ would give a power law $L \sim T^{-1/z}$.  As a result, it is natural to find either that spatial scales $L$ remain constant at small $T$ or that they diverge logarithmically.  We will see that both behaviors occur below.

Let us close with a comment on two-point functions.    At $\mu=0$, the uniqueness of the ground states and the resulting lack of TFD entanglement at small $T$ implies that (connected) correlators vanish at $T=0$.  The non-trivial ground-state entropy makes the situation different in principle for $\mu > 0$, though two-point functions with one argument in each CFT can be non-zero at $T=0$ only if each operator actually has some non-zero matrix element between two ground states of the requisite $\tilde H_1, \tilde H_2$.  The set of operators (if any) for which this occurs will depend on the detailed dynamics of the CFT.  The interesting result we will find in section \ref{sec:corr} is that, at least in the limit of large operator dimensions, this occurs precisely for operators with a certain ``extremal" ratio between their U(1) charge and conformal dimension.

\subsection{Planar Reissner-Nordstr\"om AdS}

In a holographic field theory the bulk dual of \eqref{eq:TFDmu} is straightforward to construct following \cite{Maldacena:2001kr}.  The conserved charge in the field theory will be associated with some $U(1)$ gauge field in the bulk.  We thus simply perform the bulk Euclidean path integral with boundary conditions given by the above interval $I$ and gauge field $A$.  Note that the non-trivial gauge field $A=-i\mu dt_E$ on the boundary means that the generator $\tilde H$ of bulk time-translations toward the future may be written $\tilde H  = H \pm \mu Q$, where the $+/-$ signs are respectively appropriate for systems 1 and 2 above.  Here $H$ is the generator for $\mu =0$ given by the standard expression (see e.g. e.g. \cite{Skenderis:2002wp}) for the boundary stress tensor in terms of Fefferman-Graham coefficients of the bulk metric.  See \cite{Hollands:2005ya} for a general discussion of computing time-translation generators by holographic methods for boundary conditions involving vector fields.

In the bulk semi-classical limit our path integral should be dominated by a saddle point.  We will consider cases  where this saddle point is the planar Reissner-Nordstr\"om AdS${}_{d+1}$ geometry\footnote{While the semi-classical approximation may break down in surprising ways in generic contexts involving black holes (see e.g. \cite{Mathur:2005zp,Mathur:2009hf,Almheiri:2012rt,Almheiri:2013hfa,Marolf:2013dba}), the TFD case is sufficiently special that it is plausibly free of such issues \cite{Marolf:2012xe}.}.  We expect this to be the case for $d \ge 3$ holographic field theories on Minkowski space so long as the bulk solution exhibits no instabilities associated with Schwinger pair creation~\cite{Schwinger:1951nm,Gibbons:1975kk} (see also footnote \ref{instab}).  For definiteness we consider only $d=4$ below.  Note that our path integral automatically places quantum fluctuations of bulk fields into a Hartle-Hawking-like state. Below, we use the same symbol $|\psi \rangle$ to denote the CFT state, the state of full bulk quantum gravity, and the Hartle-Hawking-like state of linearized or perturbative bulk quantum fields on the RN-AdS background.

The RN-AdS geometry solves the equations of  Einstein-Maxwell gravity with negative cosmological constant.  Taking the action to be
\begin{equation}
S = \int d^5 x \sqrt{-g} \left[\frac{1}{2\kappa^2}\left(R + \frac{12}{\ell^2}\right) - \frac{1}{4g^2} \, F^2\right],
\end{equation}
and introducing the dimensionless measure
$\gamma^2 \equiv 3g^2 \ell^2/2\kappa^2$ of the relative strengths of the gravitational and Maxwell couplings, the solutions for fixed $\mu$ may be written in terms of a scale $z_0$ that will shortly be related to the temperature $T$.
Such solutions take the form
\begin{subequations}
\label{eqs:RNAdSz}
\bea
ds^2 &= \frac{\ell^2}{z^2}\left[-f(z) \, dt^2 + \frac{dz^2}{f(z)} + dx^2_3\right], \\
A_\mu dx^\mu &= \mu(1-\zt^2) \, dt,
\eea
\end{subequations}
with
\begin{equation}
f(z) = (1-\zt^2)(1+\zt^2-\alpha^2 \zt^4), \hspace{0.5cm} \tilde z = \frac{z}{z_0}, \hspace{0.5cm} \mbox{and } \alpha^2 \equiv \frac{z_0^2 \mu^2}{\gamma^2}.
\end{equation}
The rescaling~$z_0 \rightarrow z_0/\lambda$ is equivalent to the transformation $(t,z,x) \rightarrow (\lambda t, \lambda z,\lambda x)$, $\mu \rightarrow \mu/\lambda$, so the physics depends only on the scale-invariant parameter $\alpha$, or equivalently on $\mu/T$.

\begin{figure}
\centering
\includegraphics[page=2]{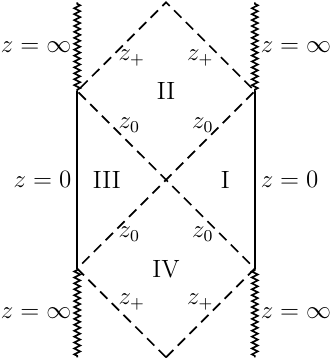}
\caption{The relevant portion of the conformal diagram of RN-AdS.  The exterior regions are~I and~III, with their boundaries at~$z = 0$.  The singularity is at~$z = \infty$, and the spacetime has inner and outer horizons at~$z = z_0$ and~$z = z_+$, respectively.  We take the imaginary part of~$t$ in regions~I-IV to be~$0$,~$\beta/4$,~$\beta/2$, and~$-\beta/4$, respectively.}
\label{fig:RNAdSconformal}
\end{figure}

The AdS boundary lies at~$z = 0$, while $z=z_0$ is a horizon with temperature
\begin{equation}
\label{eq:z0T}
T = \frac{2-\alpha^2}{2\pi z_0}.
\end{equation}
This expression can be inverted to obtain~$z_0(T,\mu)$; we will later need the small temperature behavior~$z_0 = \sqrt{2} \, \gamma/\mu + \mathcal{O}(T/\mu)$.  Note that $A_t$ vanishes at the horizon as required by regularity in a static gauge.  For nonzero~$\alpha^2 < 2$ there is also an inner horizon at~$z = z_+$, with
\begin{equation}
z_+^2 = z_0^2 \frac{1+\sqrt{1+4\alpha^2}}{2\alpha^2}.
\end{equation}
The singularity lies at $z = \infty$. The conformal diagram of maximally extended RN-AdS is shown in Figure~\ref{fig:RNAdSconformal}.

The above coordinates will be convenient despite the fact that they become singular on horizons.  The Schwarzschild-like time coordinate~$t$ should be considered to be periodic with period~$i\beta$, with~$\beta$ the inverse temperature.  Within the real Lorentz-signature solution above, we also take it to change by~$\pm i\beta/4$ whenever an outer horizon is crossed\footnote{Crossing the bifurcation surface counts as crossing two horizons and gives a change of $\pm i\beta/2$.  Upon crossing an inner horizon $t$ changes by~$i\beta_+/4$  with~$\beta_+$ the inverse temperature of the inner horizon; see e.g.~\cite{Brecher:2004gn}.  We will have no need of this in the following discussion.}.  Thus the imaginary part of $t$ determines whether a point lies in region I,II, III, or IV of the conformal diagram (Figure~\ref{fig:RNAdSconformal}). In particular, the two asymptotic regions correspond to~$\Im(t) = 0$ and~$\Im(t) = \beta/2$.  Below, it will often be useful to switch to a new radial coordinate~$w = \zt^2$, in terms of which~\eqref{eqs:RNAdSz} becomes
\begin{subequations}
\label{eqs:RNAdS}
\bea
ds^2 &= \frac{\ell^2}{z_0^2 w}\left[-f \, dt^2 + \frac{z_0^2 dw^2}{4 w f} + dx^2_3\right], \\
\label{eq:Amu}
A_\mu dx^\mu &= \mu(1-w) \, dt.
\eea
\end{subequations}

We will be most interested in the extreme limit $\alpha^2 \to 2$, where $z_+ \rightarrow z_0$ so that $f(z)$ develops a double pole at $z=z_0$. The metric with $\alpha^2=2$ has an infinite throat, as the horizon at $z=z_0$ is an infinite proper distance away in the slices of constant $t$, and a Cauchy slice in the maximally extended extremal geometry has only a single boundary. We are interested in following the entanglement between the two boundaries of the non-extremal black hole in this limit, as the length of the Einstein-Rosen bridge connecting the two asymptotic regions diverges. To this end we define $\epsilon \equiv 2- \alpha^2$ and write $f(z) = (1-\tilde{z}^2)^2(1 + 2 \tilde z^2) - \epsilon \tilde z^4 (1- \tilde{z}^2)$.

%================================================================
\section{Mutual Information}
\label{sec:mutualinfo}
%================================================================

A useful probe of the entanglement between our two Hilbert spaces is the mutual information associated with two spacetime regions, with one region in each CFT.   In the TFD context it is natural to call this thermo-mutual information (TMI) following \cite{Morrison:2012iz}, which studied the corresponding quantity for holographic theories with $\mu=0$ and $d=2$.  After a brief review, we compute TMI for general $\mu$ and $d =4$ for two strips of width $L$ located on a common bulk Killing slice, and also for regions defined by an entire CFT and a single strip of width $L$ in the other CFT.   Our main result is that
in the former case $L$ must grow as $\ln T$ near extremality in order to have non-zero TMI while it may remain finite in the latter case.

\subsection{Thermo-Mutual Information}
\label{sec:review TMI}

Recall \cite{2005PhRvA..72c2317G,2008PhRvL.100g0502W} that the mutual information between two non-overlapping regions ${\cal A}$ and ${\cal B}$ is
\begin{equation}\label{MI def}
        {\rm MI}({\cal A}: {\cal B}) =  S_{\cal A} + S_{\cal B} - S_{{\cal A} \cup {\cal B}},
\end{equation}
\noindent where $S_X = - {\rm tr} (\rho_X \log \rho_X)$ is the von Neumann entropy of the reduced density matrix $\rho_X$ describing the region $X$.  In particular, the mutual information is finite in quantum field theory as all divergences in $S_X$ are local terms at boundaries which explicitly cancel in the combination \eqref{MI def}. The mutual information is non-negative by virtue of subadditivity $S_{ {\cal A} } + S_{ {\cal B} } \geq S_{ {\cal A} \cup {\cal B} }$ for non-overlapping regions; see \cite{Lieb:1973zz,Lieb:1973cp} and also \cite{Headrick:2007km} for a holographic derivation.

The term thermo-mutual information (TMI) refers to the case where we consider a thermofield double state and the two regions are associated with different copies of the CFT. We will take ${\cal A}$, ${\cal B}$ to lie on a single Killing slice of the bulk and compute TMI holographically using the Ryu-Takayanagi
prescription \cite{Ryu:2006bv}, which instructs us to identify
\begin{equation}
\label{RT formula}
        S_X = \frac{A(\gamma_X)}{4 G_N},
\end{equation}
\noindent where $G_N$ is Newton's constant and $A(\gamma_X)$ is the area of the minimal static surface $\gamma_X$, which extends into the bulk while being i) homologous to $X$ within the surface of constant Killing time and ii) anchored on the boundary $\partial X$ of $X$. Since we will apply such recipes to bulk black holes that dominate a Euclidean path integral, this recipe can be justified using the arguments of \cite{Lewkowycz:2013nqa}.  The fact that the two regions lie at different Euclidean times $t_E =0$ and $t_E = i \beta/2$  provides no additional complications.

It is often the case that holographic TMI will vanish identically, saturating the subadditivity condition.  This occurs because the disjoint union of two minimal surfaces $\gamma_{{\cal A}_1}$ and  $\gamma_{{\cal A}_2}$ is an extremal surface anchored on the boundary of ${\cal A}_1 \cup {\cal A}_2$.  Holographic TMI vanishes when this is the minimal-area such extremal surface. In the CFT this should be considered an artifact of the large $N$ limit, though one that is unmitigated by $1/N$ corrections.  An alternative candidate for $\gamma_{{\cal A}_1 \cup {\cal A}_2}$ is a surface $\gamma_{{\cal A}_1{\cal A}_2}$ that passes through the horizon connecting the boundary of $\mathcal A_1$ to the boundary of $\mathcal A_2$.  Varying the sizes of $\mathcal A_{1,2}$ will typically result in a transition where the area of the latter surface becomes smaller than the area of the former and the TMI becomes non-zero.  The scale at which this transition occurs provides information about the degrees of freedom entangled between the two CFTs.

\subsection{Strips}
\label{TMI0}

Consider first the case where $\mathcal A_1$ and $\mathcal A_2$ are strips defined by $0 < x^1 < L$ and extending infinitely far along $x^2$ and $x^3$.  In this case the extremization problem becomes effectively one dimensional.

\begin{figure}
\centering
\includegraphics[page=3]{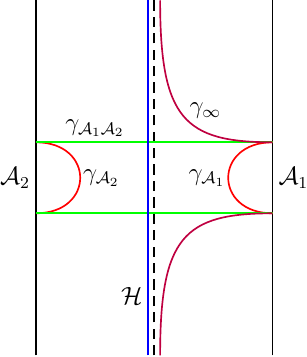}
\caption{Assorted entangling surfaces at~$t = 0$.  The boundary CFTs live on the solid lines, which show the transverse~$x^1$ direction on which we define the strips~$\cai$ and~$\caii$, which have width~$L$.  The dashed line is the bifurcation surface.  Here we show five surfaces:~$\gamma_{\cai}$ and~$\gamma_{\caii}$ correspond to the entangling surface of each strip;~$\gamma_{\cai \caii}$ runs through the bulk from one strip to the other, and can contribute to the entanglement entropy of the two strips;~$\mathcal{H}$ is a surface that runs along the horizon and corresponds to the entangling surface of the entire left CFT; and~$\gamma_\infty$ connects~$\partial \cai$ to infinity, and can contribute to the mutual information between~$\cai$ and the left CFT.}
\label{fig:minimalsurfaces}
\end{figure}

It is instructive to first review the case of finite $T = \beta^{-1}$ and $\mu = 0$~\cite{Hartman:2013qma}. Here the physics depends only on the dimensionless  combination $LT$. For our strips the connected surface $\gamma_{{\cal A}_1 {\cal A}_2}$ extends straight through the bifurcation surface, connecting~$\partial\mathcal{A}_1$ with~$\partial\mathcal{A}_2$; see Figure~\ref{fig:minimalsurfaces}.  Translational invariance implies that its area is independent of~$LT$, while the area of the disconnected surface~$\gamma_{ {\cal A}_1} \cup \gamma_{{\cal A}_2}$ vanishes as $LT \rightarrow 0$.  Thus the disconnected surface dominates~$S_{ {\cal A}_1 \cup  {\cal A}_2}$ for small $LT$ and the TMI vanishes. As we increase $LT$, the surfaces $\gamma_{\cai}$ and $\gamma_{\caii}$ reach further into the bulk but do not cross the horizon.  At large $LT$ they lie mostly along the horizon so that their areas grow linearly in $L$.  Thus there exists some critical~$L_\mathrm{strips}$ of order $T^{-1}$ such that for~$L > L_\mathrm{strips}$, the connected surface dominates in the computation of $S_{ {\cal A}_1 \cup  {\cal A}_2}$ and the TMI becomes nonzero.  This phase transition at~$L = L_\mathrm{strips}$ is sharp.  The TMI grows linearly in $L$ above the phase transition, with the leading behavior at large $L$ given by twice the thermal entropy density times the volume of either region.

We wish to investigate how this picture changes at finite chemical potential.  In particular, this allows us to study a meaningful $T \rightarrow 0$ limit with finite entanglement density between the two CFTs.  For small $\mu$ the critical $L_\mathrm{strips}$ will remain of order $T^{-1}$, but near extremality we will find that $L_\mathrm{strips}$ grows as $\ln (1/T)$.

The minimal surfaces may be found by extremizing the area functional
\begin{equation}\label{a func gen tmi}
        A = V_2 \int \frac{\ell^3}{z^3} \sqrt{\frac{dz^2}{f(z)} + d x^2} \, ,
\end{equation}
with boundaries at $z=0$ and $x^1 = 0,L$.  Here $V_2 = \int d x^2 dx^3$ is the (infinite) volume in the directions along the strip.

For the connected surface, the extremum of \eqref{a func gen tmi}  is clearly attained when $d x = 0$ in \eqref{a func gen tmi}.  Thus
\begin{equation}\label{A conn 1}
        \frac{A( \gamma_{\cai \caii} )}{4G_N} = 4 z_0V_2s \int_{\tilde z_{UV}}^1  \frac{d \tilde z}{\tilde z^{3} \sqrt{f(z)}},
\end{equation}
where $s = (1/4 G_N) \left( \ell/z_0 \right)^3 $ is the thermal entropy density and $\tilde z_{UV}$ is a dimensionless ultraviolet (UV) cutoff which we may take to zero after computing the TMI. At extremality $f$ will acquire a double pole at $\tilde z=1$ so the integral \eqref{A conn 1} diverges logarithmically. In order to extract this divergence we make the change of integration variable $u = 1-\tilde z + \epsilon$,
where $\epsilon = 2 - \alpha^2$. Expanding the integrand of \eqref{A conn 1} near $\epsilon = 0$ then gives
\begin{equation}\label{A conn app 1}
        \frac{A( \gamma_{\cai \caii} )}{4G_N} = 4 z_0 V_2 s \int_\epsilon^{u_{UV}} \frac{du}{u} \left[ \frac{1}{(u-1)^3 (2-u)\sqrt{2u^2 - 4u +3}} + O(\epsilon/u) \right],
\end{equation}
where $u_{UV} \equiv 1- \tilde z_{UV} + \epsilon$. For small $\epsilon$ and fixed $u_{UV}$, the integral \eqref{A conn app 1} is dominated by the contribution of the first term, which reduces to
\begin{equation}\label{A conn app 2}
                \frac{A( \gamma_{\cai \caii} )}{4G_N} \sim \frac{4 \gamma V_2 s}{\sqrt{6}\, \mu} \ln(\mu/T),
\end{equation}
where we used~\eqref{eq:z0T} to express~$z_0$ in terms of~$\mu$.  One may also derive \eqref{A conn app 2} by writing \eqref{A conn 1} in terms of standard elliptic integrals; see appendix \ref{app:elliptic} for details.

For the surface $\gamma_{\cai}$, \eqref{a func gen tmi} yields
\begin{equation}
                \frac{A(\gamma_{\cai})}{4G_N} = z_0 V_2 s \int_0^{L/z_0} \frac{d \tilde x}{\tilde z^3} \sqrt{ \frac{ \tilde z'^2}{f(z)} + 1},
\end{equation}
where  $\tilde x \equiv x/z_0$  and $\tilde z' = d \zt/d \tilde{x}$.  The expression for~$\gamma_{\caii}$ is of course identical. The translational symmetry in $x$ implies a conserved quantity
\begin{equation}
        \frac{1}{\tilde z^3 \sqrt{f^{-1} \tilde z'^2 + 1}} \equiv \frac{1}{\tilde z_t^3 },
\end{equation}
where $\zt_t$ is the turning point of~$\gamma_{\cai}$. Since extremal surfaces in static geometries do not penetrate horizons~\cite{Hubeny:2012ry}, we must have $\zt_t \le 1$.   The case~$\zt_t = 1$ corresponds to the surface~$\gamma_\infty$ shown in Figure~\ref{fig:minimalsurfaces}, which asymptotes to the horizon and never returns to the boundary.  For~$\zt_t < 1$ the corresponding boundary length $L$ is given by
%
%\begin{equation}\label{L disconn}
%       \frac{L}{2} = \int_0^{L/2} dx = \int_{\tilde z_{UV}}^{\tilde z_t} \frac{d \tilde z}{\tilde z'} = \int_{\tilde z_{UV}}^{\tilde z_t} \frac{d \tilde z}{f^{1/2}} %\left( \frac{\tilde z_t^{2(d-1)}}{\tilde z^{2(d-1)}} - 1\right)^{-1/2}
%\end{equation}
%
\begin{equation}
\label{L disconn}
        \frac{L}{2} = z_0 \int_0^{L/2z_0} d\tilde{x} = z_0\int_{\tilde z_{UV}}^{\tilde z_t} \frac{d \tilde z}{\tilde z'} = z_0 \int_{\tilde z_{UV}}^{\tilde z_t} \frac{d \tilde z}{\sqrt{f(z)}} \left( \frac{\tilde z_t^{6}}{\tilde z^{6}} - 1\right)^{-1/2}
\end{equation}
and the associated area is
%
%\begin{equation}\label{A disconn}
%       A_{\gamma_A}  = 2 V_2 \int_{\tilde z_{UV}}^{\tilde z_t} \frac{d \tilde z}{\tilde z'} \frac{ (f^{-1} \tilde z'^2 + 1)^{1/2}}{\tilde z^{d-1}}
%       = 2 V_2 \int_{\tilde z_{UV}}^{\tilde z_t} \frac{d \tilde z}{f^{1/2}} \left( \frac{\tilde z_t^{2(d-1)}}{\tilde z^{2(d-1)}} - 1\right)^{-1/2}
%       \frac{\tilde z_t^{d-1}}{\tilde z^{2(d-1)}}
%\end{equation}
%
\begin{equation}
\label{A disconn}
        \frac{A(\gamma_{\cai})}{4G_N}  = 2 z_0 V_2 s \int_{\tilde z_{UV}}^{\tilde z_t} \frac{d \tilde z}{\tilde z'} \frac{ \sqrt{f^{-1} \tilde z'^2 + 1}}{\tilde z^{3}}
        = 2  z_0 V_2 s \int_{\tilde z_{UV}}^{\tilde z_t} \frac{d \tilde z}{\sqrt{f(z)}} \left( \frac{\tilde z_t^{6}}{\tilde z^{6}} - 1\right)^{-1/2}
        \frac{\tilde z_t^{3}}{\tilde z^{6}}.
\end{equation}

Since \eqref{A conn app 1} grows as $T \rightarrow 0$, one can obtain nonzero TMI at small $T$ only when \eqref{A disconn} is similarly large.  This occurs when  $L$ is large and
$\tilde z_t \approx 1$. From \eqref{L disconn} and \eqref{A disconn} we find in this regime that
$A(\gamma_{\cai}) = L V_2 s + \mathcal{O}(1)$,
describing the extensive thermal entanglement expected for large $L$.
Comparison with~\eqref{A conn app 2} implies that for small~$T$, the transition to TMI~$> 0$ occurs at
\begin{equation}
        L_\mathrm{strips} = \frac{\gamma}{\sqrt{6}\, \mu} \ln (\mu/T)  + \mathcal{O}(1).
\end{equation}
As advertised, an infinite growth of the entangling regions is required to obtain a non-vanishing TMI near extremality.

\subsection{A strip and an entire CFT}
\label{sec:s+CFT}

\begin{figure}[t]
\centering
\includegraphics[page=4]{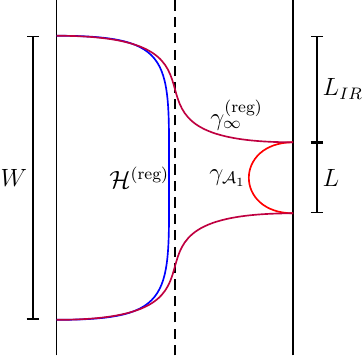}
\caption{Surfaces relevant to the IR regularization used to compute~$\mathrm{TMI}(\cai:\mathrm{CFT}_2)$.}
\label{fig:regsurfaces}
\end{figure}

In contrast to the above, let us now consider the mutual information between a finite strip~$\cai$ in one CFT and the entire second CFT.  The calculations are similar to those just performed.  Defining a surface~$\gamma_\infty$ that ends on~$\partial \cai$ and extends to infinity on the other boundary (as shown in Figure~\ref{fig:minimalsurfaces}), we have
\begin{equation}
\label{s+CFT}
\mathrm{TMI}(\cai:\mathrm{CFT}_2) = \mathrm{max}\left\{0, sV_3 + \frac{A(\gamma_{\cai})  -  A(\gamma_\infty)}{4 G_N} \right\},
\end{equation}
where $V_3$ is the spatial volume of the CFT.  While~$s V_3$ and~$A(\gamma_\infty)$ are both~IR divergent, one can easily show that these divergences cancel.  To do so, first consider the mutual information between the strip~$\cai$ of width~$L$ in one CFT and a strip of width~$W$ in the other, with~$W$ large relative to any other scale; the relevant entangling surfaces ${\cal H}^\mathrm{(reg)}$ and $\gamma_\infty^\mathrm{(reg)}$ are shown in Figure~\ref{fig:regsurfaces}.  The desired result is obtained in the limit~$W \rightarrow \infty$, so that~$W$ serves as an IR regulator.  The length~$L_\mathrm{IR}$ of one of the regulated surfaces~$\gamma_\infty^\mathrm{(reg)}$ (see Figure \ref{fig:regsurfaces}) is given by~\eqref{L disconn} with~$\zt_t = \zt_\mathrm{IR} \equiv 1 + \delta$, with~$\delta$ small and positive\footnote{In fact, because~$\zt_\mathrm{IR} > 1$, the upper bound of the integral in~\eqref{L disconn} should be set to~1.}.  At large $W$ the entropy of the strip of width~$W$ will approach~$s W + 2S_0 = s(L + 2L_\mathrm{IR}) + 2 S_0$, where the $W$-independent correction $S_0$ is associated with the part of ${\cal H}^\mathrm{(reg)}$ that stretches from the horizon to the left boundary.  Since this same correction appears in the area of $\gamma_\infty^\mathrm{(reg)}$ we find
\begin{equation}
\label{diff}
s V_3 - \frac{A(\gamma_\infty)}{4G_N} = \frac{V_2 \ell^3}{G_N z_0^2} \left[ \frac{L}{4 z_0} - \int_0^1 \frac{1}{\zt^3}\sqrt{\frac{1-\zt^6}{f(z)}} \, d\zt\right] + \mathcal{O}(\delta).
\end{equation}
The divergences have canceled as promised.

\begin{figure}[t]
\centering
\includegraphics[width=0.5\textwidth]{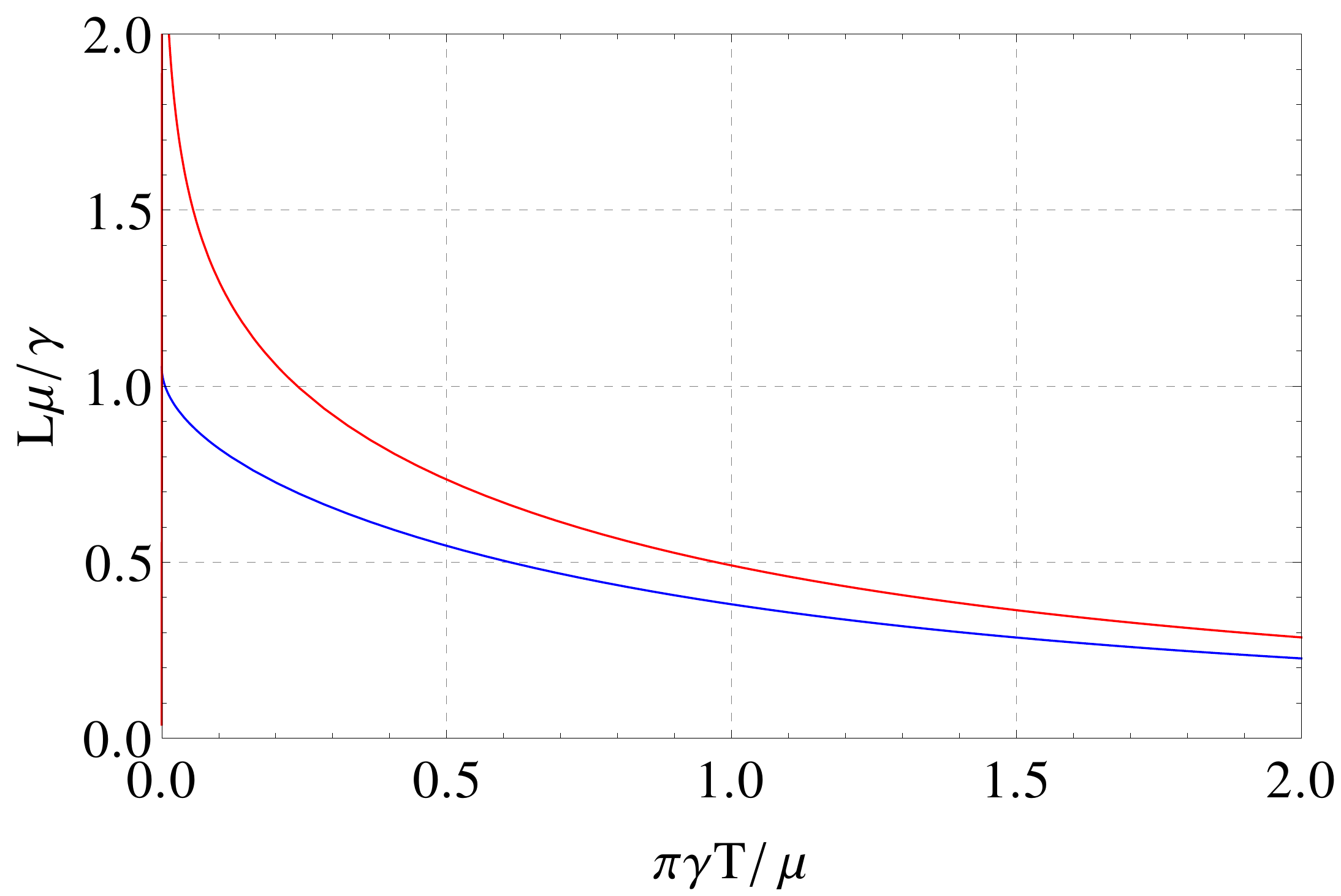}
\caption{The critical lengths~$L_\mathrm{strips}$ (upper curve, red) and~$L_{s+\mathrm{CFT}}$ (lower curve, blue) as functions of temperature.  $L_\mathrm{strips}$ diverges logarithmically at small~$T$, whereas~$L_{s+\mathrm{CFT}}$ approaches a constant value~$\approx 1.05 \gamma/\mu$. }
\label{fig:TMI}
\end{figure}

Since the UV-regularized value of $A(\gamma_{{\cal A}_1})$ is finite and monotonically increasing with $L$, we see that the regulator-independent quantity
\begin{equation}
\label{3terms}
sV_3 + \frac{A(\gamma_{\cai})  -  A(\gamma_\infty)}{4 G_N}
\end{equation}
relevant to \eqref{s+CFT} grows linearly at large $L$ and diverges to $-\infty$ as $L \rightarrow 0$.  Thus there is a critical length $L_{s+\mathrm{CFT}}(T/\mu)$ at which \eqref{s+CFT} becomes non-zero, given by requiring  \eqref{3terms} to vanish.  The results are shown in Figure~\ref{fig:TMI}, and we find numerically that $L_{s+\mathrm{CFT}}|_{T/\mu=0} \approx 1.05 \, \gamma/\mu$.

\begin{figure}[t]
\centering
\subfloat[\label{subfig:Acomplement1}]{
\includegraphics[page=5]{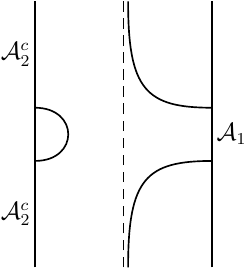}
}
\hspace{0.7cm}
\subfloat[\label{subfig:Acomplement2}]{
\includegraphics[page=6]{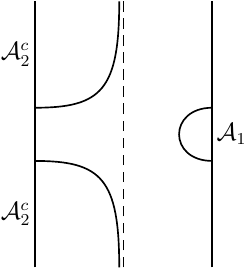}
}
\hspace{0.7cm}
\subfloat[\label{subfig:Acomplement3}]{
\includegraphics[page=7]{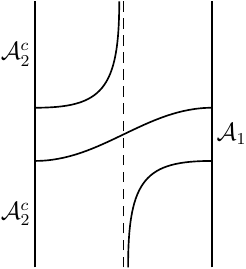}
}
\caption{Surfaces relevant to the entanglement entropy~$S_{{\cal A}_1 \cup {\cal A}_2^c}$.  Note that the surfaces shown in~(a) and~(b) are reflections of one another across the horizon.  Another surface (not shown) is related to~(c) by a reflection in the vertical direction; i.e., across~$x_1 = 0$.}
\label{fig:Acomplement}
\end{figure}

The contrast between $L_\mathrm{strips}$ and $L_{s+\mathrm{CFT}}$ is striking.  A further interesting result is obtained by considering  again two strips ${\cal A}_1$ and ${\cal A}_2$ of width $L$, but this time considering the TMI between ${\cal A}_1$ and ${\cal A}_2^c$, the complement of ${\cal A}_2$ in CFT${}_2$. The possible bulk surfaces for computing $S_{{\cal A}_1 \cup {\cal A}_2^c}$ look much like those studied above (see Figure~\ref{fig:Acomplement}).\footnote{They are also similar to those that might be used to compute $\mathrm{MI}({\cal A}_1: {\cal A}_1^c)$, with both regions in the same CFT.  But this is strictly infinite.  The UV divergences do not cancel in \eqref{MI def} when the regions overlap.}  If we take the limit of small $T$ at fixed $L,\mu$, the surface that connects the two boundaries  in Figure~\ref{subfig:Acomplement3} will have divergent area, so it cannot dominate the others.  But the two remaining surfaces allowed by homology (point (i) below \eqref{RT formula}) are related by reflection through the horizon and so have equal areas.  Since one surface is just $\gamma_{{\cal A}_1} \cup \gamma_{{\cal A}_2^c}$, we find $\rm{TMI}({\cal A}_1: {\cal A}_2^c) =0$.  Thus we have the remarkable result that for finite $L$ the TMI between ${\cal A}_1$ and either ${\cal A}_2$ or its complement vanish in the limit of small $T$, but the TMI between ${\cal A}_1$ and the entire other copy of the CFT can remain non-zero.

This result is not readily accommodated by the localized-quasiparticle picture of TFD entanglement (see \cite{Hartman:2013qma}, following \cite{Calabrese:2009qy} in the time-dependent case; see also \cite{Liu:2013iza,Liu:2013qca} for other features that indicate shortcomings of this model).   A quasiparticle picture might suggest that any entanglement between
${\cal A}_1$ and CFT${}_2$ should be visible even if we separate CFT${}_2$ into ${\cal A}_2$ and ${\cal A}_2^c$. That is, a localized-quasiparticle picture would lead us to expect that at least approximately
\begin{equation}
\label{quasi}
\rm{TMI}({\cal A}_1: CFT{}_2) = \rm{TMI}({\cal A}_1: {\cal A}_2) + \rm{TMI}({\cal A}_1: {\cal A}_2^c).
\end{equation}
This expectation is badly violated in our case at small $T$, as the former TMI is non-zero, but both of the terms on the other side vanish. This just a particularly striking example of a general failure of \eqref{quasi}.  In the holographic context this is because the mutual information between some regions ${\cal A}$ and ${\cal B}$ on the one hand, and mutual information between ${\cal A}$ and subregions ${\cal B}_1$, ${\cal B}_2$ on the other, will involve different surfaces, and there is no reason to expect their areas to be related in such a way as to make \eqref{quasi} valid even approximately. While we offer no better model, it would be interesting to reflect further on what such a model might require, and perhaps to connect it with information-theoretic phenomena such as information locking \cite{2004Divincenzo,2004Hayden}.

%================================================================
\section{Charged Correlators}
\label{sec:corr}
%================================================================

We now turn to demonstrating that our TFD entanglement can be seen near extremality in the correlation functions of charged operators, and to extracting features of this entanglement.  We work in the approximation of large operator dimension, where the computation of dual bulk two-point correlators amounts to finding appropriate spacelike world lines extending from one boundary to the other. Though this approximation breaks down in certain interesting regimes, it nevertheless provides many useful results.  We begin with a detailed discussion of general charged correlators in non-extreme black hole backgrounds and specialize to the near-extreme case only in section \ref{sec:full}.

\subsection{Holographic Two-Point Functions in the worldline approximation}
\label{sec:TFD}

Let us briefly review the connection between CFT two-point functions and bulk worldlines.  In order to probe the entanglement inherent in our charged TFD state, we will be particularly interested in two-point correlation functions involving an operator $\mathcal{O}_1$ acting on one CFT in our TFD and an operator $\mathcal{O}_2$ acting on the other.   The construction of \eqref{eq:TFDmu} suggests that we take $\mathcal{O}_2$ to be the time-reverse of $\mathcal{O}_1$.  For typical complex scalar fields, this amounts to taking the adjoint: $\mathcal{O}_2 = \mathcal{O}^\dagger_1$.

Recall now that CFT scalar operators~$\mathcal{O}$ with conformal dimension~$\Delta$ are holographically related to bulk scalar fields~$\phi$ with mass~$m$ via
\begin{equation}
\Delta = \frac{d}{2} + \sqrt{\frac{d^2}{4} + m^2 \ell^2},
\end{equation}
with~$\ell$ the AdS radius and~$d$ the boundary dimension\footnote{For~$m^2$ near the Breitenlohner-Freedman bound~\cite{Breitenlohner:1982bm}, the bulk field may satisfy alternate boundary conditions in which case we have
$\Delta = d/2 - \sqrt{d^2/4 + m^2 \ell^2}$ \cite{Klebanov:1999tb}.  But this is not relevant for us since we take $m^2$ large and positive.}. At leading order in the bulk semi-classical limit, the CFT two-point function
$G_{12} =\left\langle \psi \middle| \mathcal{O}_2(x_2) \mathcal{O}_1(x_1) \middle | \psi \right\rangle$ is dual to a certain rescaled limit of the bulk two-point function $\mathcal{G}(x_1,z_1,x_2,z_2) = \langle \psi |\phi^\dagger(x_2,z_2) \phi(x_1,z_1)| \psi \rangle$ as $z_1,z_2 \rightarrow 0$.  In the former expression $|\psi \rangle$ represents the CFT charged TFD \eqref{eq:TFDmu}, while in the latter expression $\phi$ is an otherwise-free (i.e., linearized) charged field on RN-AdS and $|\psi \rangle$ is the associated Hartle-Hawking state defined by our Euclidean path integral.  Here we consider Wightman two-point functions for definiteness, though in both the CFT and bulk our primary interest will be in two-point functions of commuting operators so that the Wightmann and time-ordered two-point functions coincide.

In the limit of large~$m$ the bulk two-point function can be studied using the WKB approximation. Since our bulk quantum state was constructed from a Euclidean path integral, for neutral scalars this reduces to the familiar result $\mathcal{G} \sim e^{im\Delta \tau}$, where $\Delta \tau$ is the proper time that elapses along the geodesic connecting $(x_1,z_2)$ to $(x_2,z_2)$; see e.g. \cite{Louko:2000tp,Festuccia:2005pi}.  When the geodesic is spacelike it is more natural to write $\mathcal{G}(x_1,x_2) \sim e^{-mL}$, where $L$ is the proper length.  Taking $z_1,z_2\rightarrow 0$ and performing the above-mentioned rescaling gives $G_{12} \sim e^{-mL_\mathrm{reg}}$, where $L_\mathrm{reg}$ is an appropriately regulated version of the geodesic length $L$.

It is straightforward to generalize this result to charged operators. If the operator~$\mathcal{O}$ is charged under our global $U(1)$ symmetry, then the dual bulk operator $\phi$ is charged under the associated bulk Maxwell field.  So before integrating over paths, the proper time $\Delta \tau$ above should be replaced with the action~$S$ of a charged particle.  The relevant saddle points are then extrema of $S$, which are generally not geodesics, and the Wightman function becomes~$\mathcal{G} \sim e^{im S}$.  Due to our interest in spacelike separated points at opposite boundaries of the bulk, we write ~$G_{12} \sim e^{-mI}$ with
\begin{equation}
\label{eq:action}
I = -iS_\mathrm{reg} = \int \left(\sqrt{g_{\mu\nu} u^\mu u^\nu} - \frac{i q}{m} A_\mu u^\mu\right) \, d\lambda + I_\mathrm{ct}, \ \ \quad I_\mathrm{ct} = - \ell \ln \left( \frac{4}{w_{UV}} \right),
\end{equation}
where~$q$ and~$m$ are the charge and mass of~$\phi$, $A_\mu$ is the Maxwell field in the bulk solution. In \eqref{eq:action}, the $I_\mathrm{ct}$ is the appropriate counter-term which that makes the result finite for $z_1 = z_2 = 0$ and thus enacts the above-mentioned rescaling of bulk correlators near the boundaries.  This $I_\mathrm{ct}$ is independent of $q$ and thus identical to the standard counter-term for neutral particles; i.e., it is associated with the divergent length of geodesics near the boundaries.   As usual, we understand \eqref{eq:action} to be defined by first evaluating both $I_\mathrm{ct}$ and the bulk term with UV cutoffs and then taking the limit where the cutoffs are removed.  The detailed justification of the bulk Euclidean action \eqref{eq:action} is provided in Appendix~\ref{app:action}, in part because this expression corrects certain errors in the literature.

We emphasize that, in our background, the expression \eqref{eq:action} computes $\mathcal{G}(x_1,x_2)$ with time dependence generated by $\tilde H_1 = H + \mu Q$, $\tilde H_2 = H-\mu Q$.  Recalling that the limit $T \rightarrow 0$ with fixed $\mu$ restricts \eqref{eq:TFDmu} to terms with a unique value of $\tilde E_1 = E_1 +\mu Q_1$, one sees that either time-translation of \eqref{eq:TFDmu} changes the $T=0$ wavefunction only by an overall phase.  So in this limit $G_{12}$ should become time-independent in either argument.

\subsection{Equations of Motion}

The spacelike world lines we seek extremize the action~\eqref{eq:action}.  We take both end points to have the same spatial coordinates $\vec x$ in the directions along the planar black hole.    Parity symmetry and momentum conservation then guarantee that $\vec x$ is constant along our world line. Without loss of generality we henceforth set $\vec x =0$.

Thus our curves will have tangent vectors~$u^\mu = (\dot{t}, \dot{w}, 0, 0, 0)$.  We may use the Killing field~$\partial_t$ to introduce a conserved (Euclidean) energy~$E_E$
\begin{equation}
\label{consE}
\left(\partial_t\right)^\mu u_\mu = iE_E + i \frac{q}{m} \left(\partial_t\right)^\mu A_\mu.
\end{equation}
Together with the normalization condition~$u^\mu u_\mu = 1$, the world lines must satisfy the equations of motion
\begin{subequations}
\bea
\label{eq:tdot}
\dot{t} &= -\frac{z_0}{\ell} \, i w \frac{\E + \Q(1-w)}{f(w)}, \\
\label{eq:wdot}
\dot{w}^2 &= \left(\frac{2}{\ell}\right)^2 \, w^2 g(w),
\eea
\end{subequations}
where~$\Q = z_0q\mu/m\ell$,~$\E = z_0 E_E/\ell$, and
\begin{equation}
\label{eq:fgdiff}
g(w) = f(w) - w\left(\E + \Q(1-w)\right)^2.
\end{equation}
We also define~$\qt = qg/m\kappa$ as a dimensionless measure of the charge-to-mass ratio of~$\phi$.

Recall that we consider theories for which RNAdS${}_5$ remains stable close to extremality.  This in turn restricts the possible scalar fields that can exist in the bulk.  In particular, we wish to avoid any Schwinger pair creation instability (see again footnote \ref{instab}) \cite{Schwinger:1951nm,Gibbons:1975kk}.  In the worldline approximation, this instability arises when electrostatic repulsion of the associated particles from the black hole overwhelms the gravitational attraction.  This issue is readily analyzed by studying the potential $V(w)$ which controls motion of quasi-static (i.e., non-relativistic) timelike worldlines.  For each black hole (with, say, positive charge), there is some critical positive $\qt_\mathrm{crit}$ and which $V(w)$ develops a minimum outside the horizon.  One finds $\qt_\mathrm{crit} > 1$ for all nonextreme black holes and $\qt_\mathrm{crit} =1$ for extreme black holes.  For simplicity, we therefore restrict discussion below to the case $\qt \le 1$ unless otherwise noted.

We are interested in world lines running from $(t,w) = (t_b,0)$ to $(t,w) = (-t_b + i \beta/2,0)$.  For fixed $t_b$ there will generally be a finite set of solutions to \eqref{eq:tdot}, \eqref{eq:wdot} distinguished by their values of $\E.$   These values are generally complex, though \eqref{eq:tdot} implies that one may find solutions with imaginary $t_b$ having real $\E$.  Since there are multiple solutions, the full set of solutions for all complex $t_b$ may be associated with a Riemann surface $\E(t_b)$.  While we will focus on the curves defined by taking $t_b$ real, the $i$ that accompanies the Maxwell term in \eqref{eq:action} makes it particularly natural to analytically continue to complex parameters.

It will also be useful to characterize solutions by their turning points $w_t$ in the complex $w$-plane.  Such turning points are defined by noting that \eqref{eq:wdot} is invariant under changing the sign of the affine parameter $\lambda$ along the worldline while holding $\E, \Q, \mu$  fixed.  Thus one obtains the same curve $w(\lambda)$ whether one integrates \eqref{eq:wdot} starting from $w=0$ at the right boundary or from the left, and each solution of interest has a $Z_2$ symmetry mapping $w(\lambda)\rightarrow w(-\lambda)$ with a corresponding action on $t(\lambda)$.  The turning point is just the value of $w$ at the fixed point, $w_t \equiv w(\lambda=0)$.

\begin{figure}[t]
\centering
\includegraphics[width=0.4\linewidth]{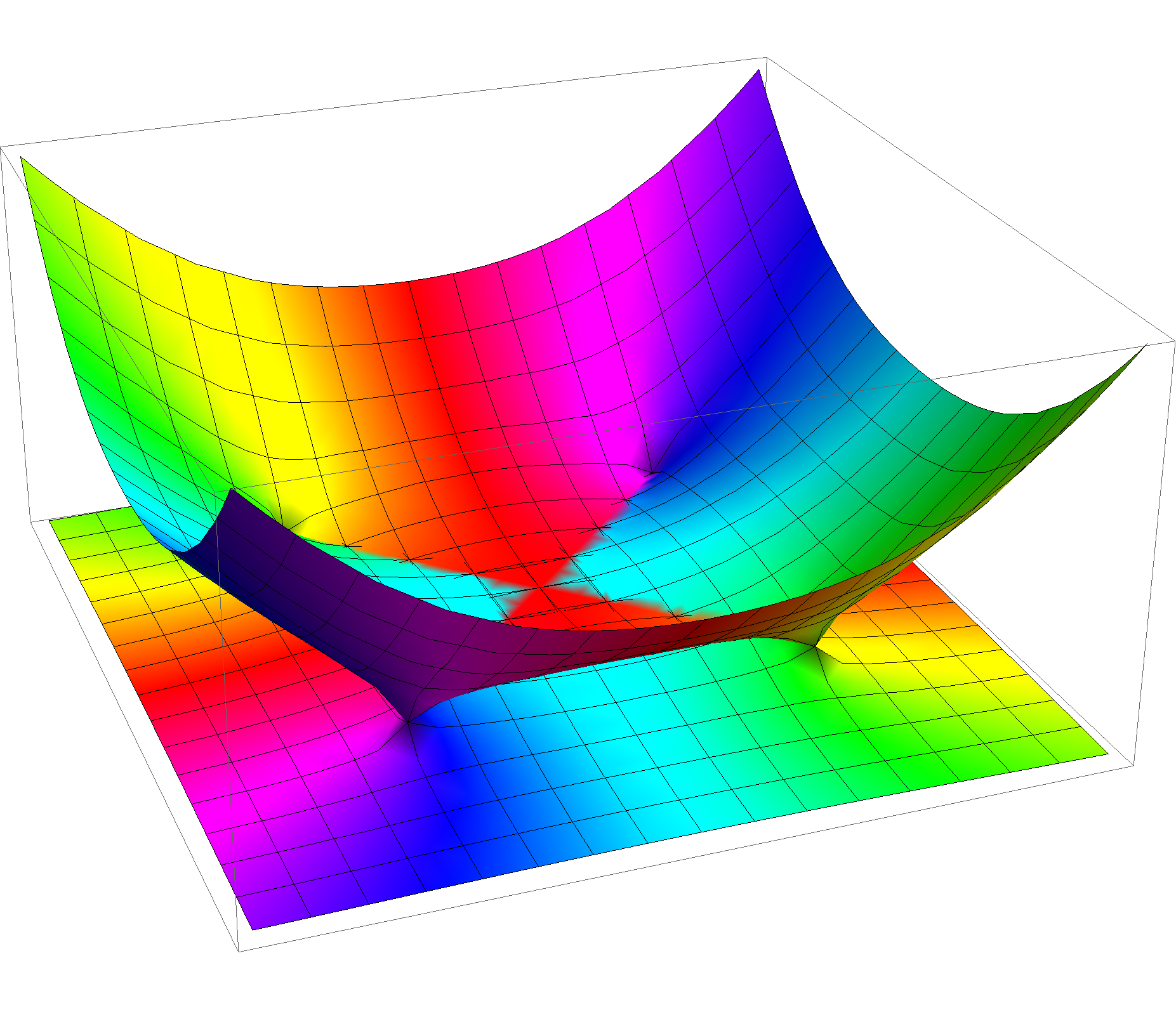}
\hspace{0.75cm}
\includegraphics[width=0.4\textwidth]{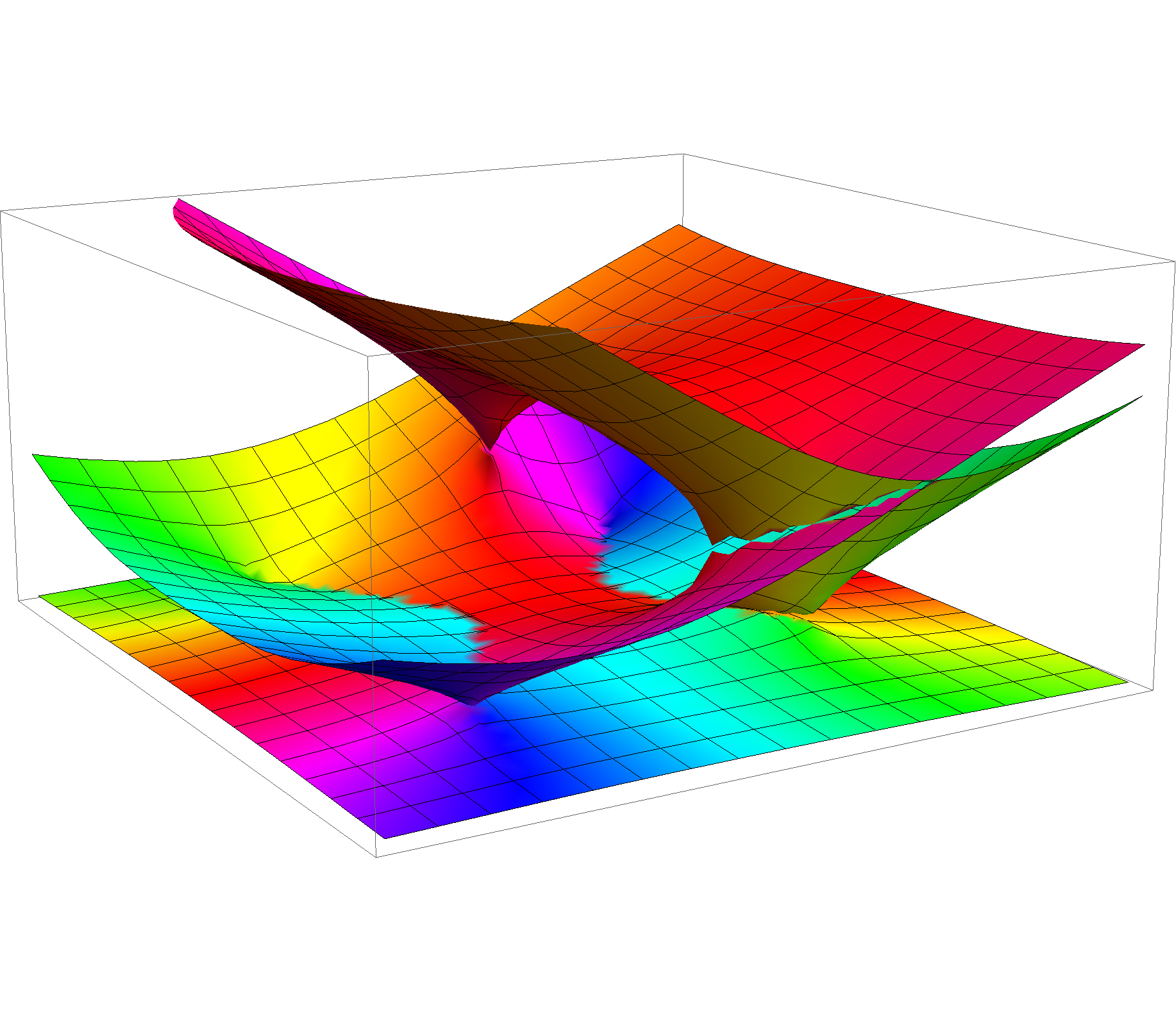}
\\
\includegraphics[width=0.4\textwidth]{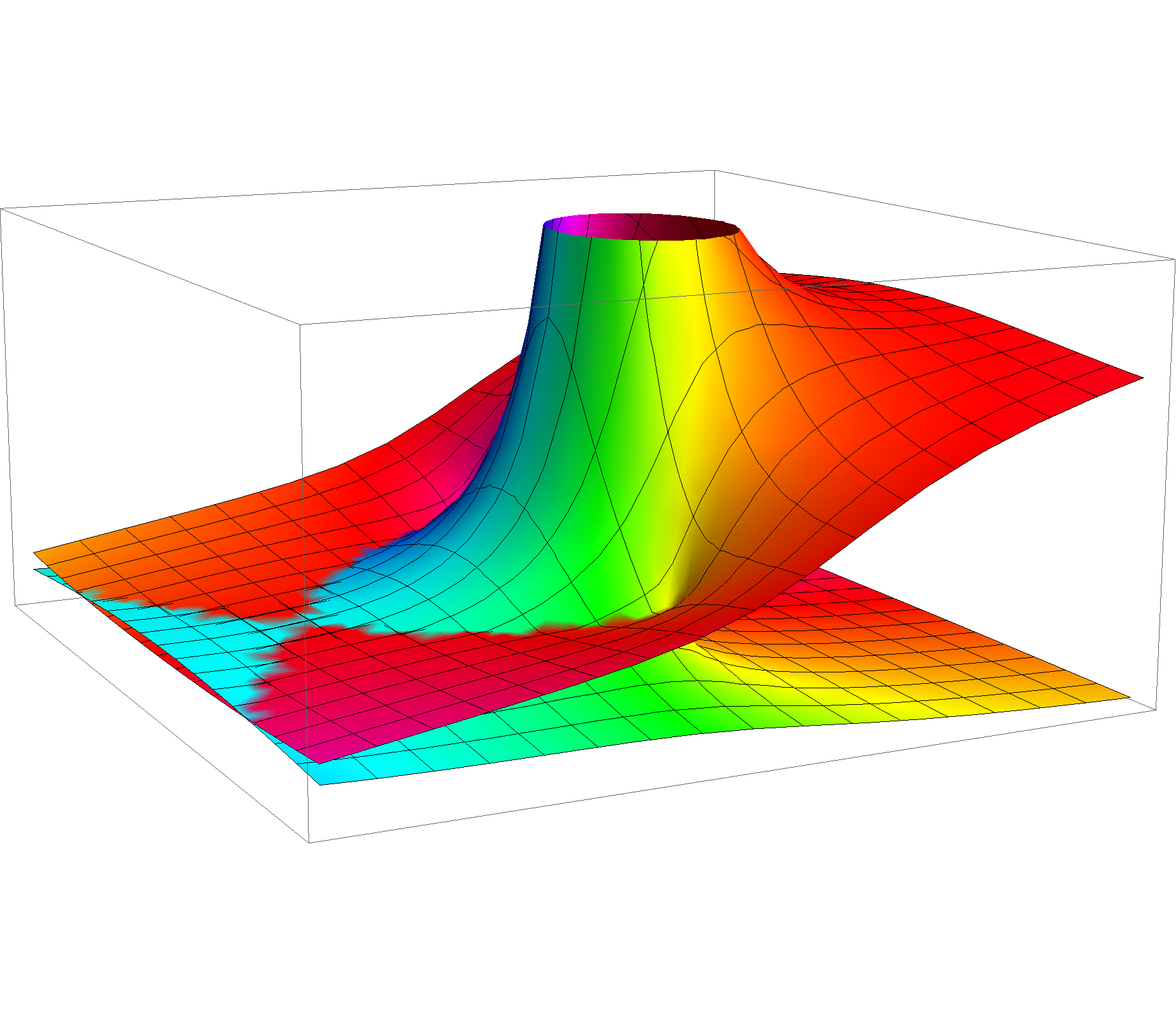}
\hspace{0.75cm}
\includegraphics[width=0.4\textwidth]{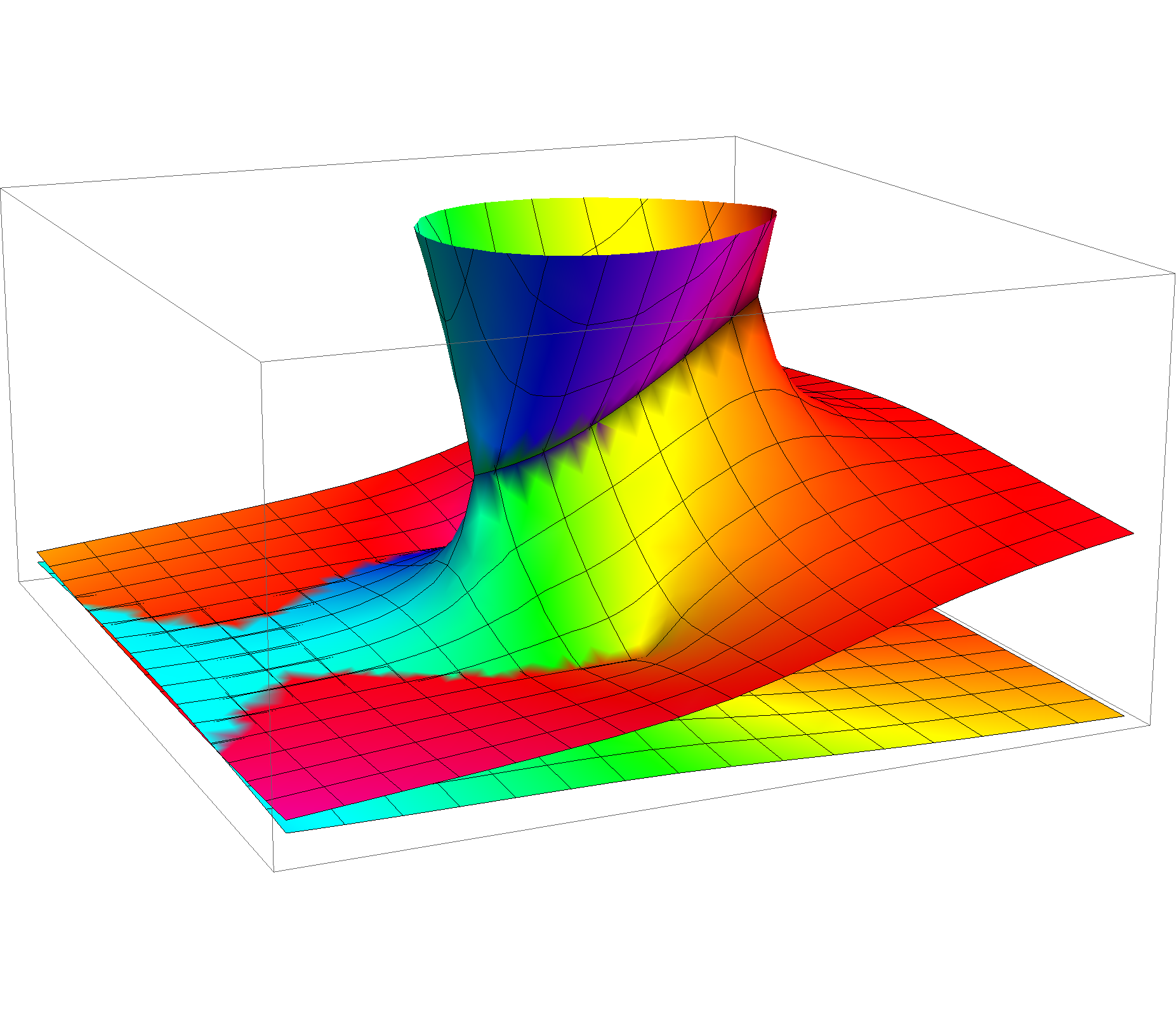}
\caption{Typical Riemann surfaces $w_t(\E)$ defined by setting $g(w_t) = 0$ over the complex~$\E$ plane.  The top left figure displays the Schwarzschild case~$\alpha = 0$; the next three show~$\alpha = 0.5$, with~$\qt = 0.2$,~$\sqrt{2/3}$, and~$0.9$, from left to right and top to bottom.  The height of the sheets corresponds to~$|w_t|$, while the hue represents the phase of $w_t$ (with red and turquoise corresponding to positive and negative real~$w_t$, respectively).  Along the real axis (parallel to the common plane of symmetry in each figure), the turning point~$w_t$ corresponds to the smallest positive real root, which is then analytically continued to the rest of the complex plane.  Thus the principal branch of $w_t$ becomes the the lowermost sheet in each figure at large real $\E$. }
\label{fig:wbranches}
\end{figure}

Note that $\dot{w}$ must vanish at this fixed point,  so that $w_t$ and $\E$ satisfy a relation given by setting $g(w_t) =0$.  Since $g$ is a cubic polynomial in $w$, one may take this to define a three-sheeted Riemann surface $w_t(\E)$ (see e.g. Figure~\ref{fig:wbranches}) with branch points corresponding in general to double roots of $g$.    This structure will play an important role below. For real $\alpha$ our $g$ has a triple root only for the special case $\E=0$, $\qt=1$ at extremality ($\alpha = \sqrt{2})$, where the root lies at the horizon ($w = 1$).

In general, the term on the right-hand side of~\eqref{eq:wdot} acts as an effective (possibly complex) Newtonian potential for $w$.  This understanding allows one to write the total elapsed time $\Delta t = -2 t_b+ i \beta/2$ and the action $I$ along any solution in the form
\begin{subequations}
\label{eqs:DeltatI}
\bea
\label{eq:Deltat}
\Delta t &= -\frac{i z_0}{2} \oint \frac{\E + \Q(1-w)}{f(w) \sqrt{g(w)}} \, dw, \\
\label{eq:I}
I &= \frac{\ell}{2} \oint \frac{f(w) - \Q \, w (1-w)\left(\E + \Q(1-w)\right)}{wf(w) \sqrt{g(w)}} \, dw + I_\mathrm{ct},
\eea
\end{subequations}
where the integral is over the contour in the complex $w$-plane defined by our worldline. Expressions \eqref{eq:Deltat}, \eqref{eq:I} use the prescription of~\cite{Hartman:2013qma} for integrating through zeros of $f$ so that crossing any horizon adds $\pm i \beta/4$ to $\Delta t$ as desired.
As a technical note we mention that by writing
\begin{equation}
g(w) = \left(\alpha^2 - \Q^2\right)(w_1 - w)(w_2 - w)(w - w_3)
\end{equation}
with~$w_t = w_1$, the expressions~\eqref{eqs:DeltatI} can be evaluated explicitly in terms of standard elliptic integrals.  These expressions are functions of~$w_1$,~$w_2$,~$w_3$,~$\alpha$,~$\Q$, and~$\E$ given in Appendix~\ref{app:elliptic} and are useful for various asymptotic expansions.

In order to obtain a unique value from  \eqref{eqs:DeltatI} we must specify $\sqrt{g}$ along this worldline which, as noted above, will necessarily run through a root $w_t$ of $g$.  The correct prescription is determined by taking $\sqrt{g}$ to be continuous along the worldline and requiring the above reflection symmetry $\lambda \rightarrow -\lambda$ to change the sign of $\sqrt{g}$; roughly speaking, the sign of $\sqrt{g}$ changes when one passes through the turning point.
The remaining sign ambiguity in $\Delta t$ is fixed by the sign of $dw/dt$ at any point along our worldline, while the sign of the ambiguity in $I$ is fixed by the condition that the divergence at $w=0$ is canceled by $I_\mathrm{ct}$.  In particular, although there are two solutions for given $\E,w_t$, the action $I$ takes identical values on both.

One would like to think of \eqref{eq:Deltat} and \eqref{eq:I} as defining $I$ as a function of $t_b$.  But again the multiple geodesics for each $t_b$ mean that $I(t_b)$ is actually a multi-sheeted Riemann surface.  A useful way to deal with this complication is to parametrize both $I$ and $t_b$ by the energy $\E$.  While the resulting $I(\E)$ and $\Delta t(\E)$ are again multi-sheeted Riemann surfaces, their structure is closely related to the physics of quasi-normal modes.  We review this connection in section \ref{sec:latetime} below and use it to extract the most relevant features.

An additional simplifying feature of this perspective is that all branch points in $I(\E)$ coincide with those of $w_t(\E)$, and thus with $w_t$ being a double root of $g$. This follows from the above observation that $I$ is uniquely determined once both $\E,w_t$ are specified.  Specializing for the moment to non-extreme black holes, we see from \eqref{eq:fgdiff} that $f$ cannot vanish where $g$ has a double root\footnote{Note that $g(w=0) = 1 \neq 0$.  Thus from \eqref{eq:fgdiff} $f$ and $g$ can vanish simultaneously at $w_0$ only when $\E + \Q(1-w_0) =0$.  But this forces the second term in \eqref{eq:fgdiff} to have a double root.  So if $w_0$ is a double root of $g$, would also be a double root of $f$.  And when $\alpha$ is real $f$ can have a double root only at extremality.}.  As a result,  the relation \eqref{eq:fgdiff} ensures that no further factors vanish at $w_t$ in either the numerator or denominator of \eqref{eq:I} and that double zeros of $g$ give logarithmic branch points.   We see that $\Delta t$ also diverges logarithmically at branch points of $w_t$ and that the only additional branch points in $\Delta t(\E)$ are those associated with the overall choice of sign.  These play only a very minor role and are not associated with divergences unless they coincide with those above.  Thus the branch points of $w_t$ are directly associated with the late time limit $\Delta t \rightarrow \infty$.

As a final note, we mention that equations~\eqref{eqs:DeltatI} and the associated boundary conditions are invariant under the transformation~$(t_b, \Q, \E) \rightarrow (-t_b, -\Q, -\E)$ and also under~$(t_b, \Q, \E) \rightarrow (-t_b, \Q, \bar{\E})$ where the overbar denotes complex conjugation.  Without loss of generality we may thus restrict our analysis to~$t_b \geq 0$ and~$\Q \geq 0$.  We may also restrict ourselves to~$\mu \geq 0$ (and thus $\qt \ge 0$), since the equations are also invariant under~$(\mu, q) \rightarrow (-\mu, -q)$.

\subsection{The Late-Time Limit and quasinormal modes}
\label{sec:latetime}

We noted above that our problem is associated with multi-sheeted Riemann surfaces $I(\E)$,  $t_b(\E)$, $w_t(\E)$ for which the interesting branch points occur when $w_t$ is a double zero of $g(w)$. Furthermore, these are precisely the points associated with late-time limits.  We now take a moment to understand the structure of these branch points in detail and to more carefully review the connection with late times.  In particular, we relate the associated branch cuts with families of quasi-normal modes drawing heavily from \cite{Festuccia:2005pi} and \cite{Hartman:2013qma}.

\begin{figure}
\centering
\includegraphics[page=8]{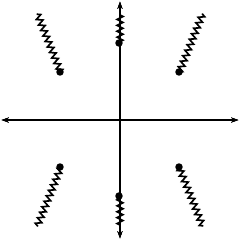}
\hspace{1cm}
\includegraphics[page=9]{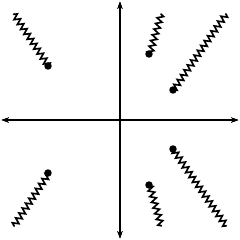}
\hspace{1cm}
\includegraphics[page=10]{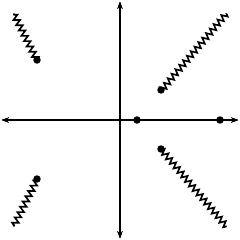}
\caption{Roots of the discriminant of~$g(w)$ in the complex~$\E$ plane are shown as dots for representative choices of the parameters~$\alpha$,~$\qt$.  From left to right, we take~$\qt = 0$,~$\qt < \sqrt{2/3}$, and~$\qt > \sqrt{2/3}$.
These indicate possible branch points for $w_t(\E), I(\E), \Delta t(\E)$, with the actual branch structure of
$I(\E), \Delta t(\E)$ being determined by that of $w_t(\E)$.  On sheets where $\E_c$ is indeed a branch point, both $I$ and $\Delta t$ diverge logarithmically.  As these are the only locations where $\Delta t$ can diverge, they serve as endpoints for all curves $\E(t_b)$. The jagged lines are rough guesses for the locations of the branch cuts that define the principal sheet of~$w_t(\E)$, $I(\E)$ and~$\Delta t(\E)$, and should correspond to lines of poles in frequency space correlators for operators of large-but-finite conformal dimension.
}
\label{fig:Estar}
\end{figure}

Note that double roots of $g$ can arise only at special values $\E_c$ of $\E$ at which the discriminant of $g$ vanishes. This discriminant is a sixth order polynomial in~$\E$ and, while its explicit form is unilluminating,  we plot the associated six roots~$\E_c$ in the complex~$\E$ plane for representative choices of the parameters~$\alpha$,~$\qt$ in Figure~\ref{fig:Estar}.  Any curve~$\E(t_b)$ must approach one of these points as $t_b \rightarrow \pm \infty$. There is a special $\mu$-independent value~$\qt = \sqrt{2/3}$ at which two of the~$\E_c$ merge on the real axis and disappear.  This corresponds to a degenerate case where~$g(w)$ becomes quadratic in~$w$, so at this value there are only four~$\E_c$.  For~$\qt < \sqrt{2/3}$ no~$\E_c$ lie on the real axis, while for~$\qt > \sqrt{2/3}$ two of the~$\E_c$ always lie on the real axis.

As described in \cite{Festuccia:2005pi}, there is a very physical relationship between the critical energies~$\E_c$ and the quasi-normal modes (QNMs) of the scalar field probe. In general, the long-time behavior of two-point functions of fields on a black hole background is dominated by the lowest QNM~$\omega_c$. This is usually used as an approximation for the two-point function in one static region outside the black hole, but continuing one of the points to the other asymptotic region via $t \to -t + i \beta/2$ one sees that it also provides an approximation to our Wightman function:
\begin{equation}
\label{eq:Iomega}
G_{12}(t_b) \sim e^{-2i\omega_c t_b} \Rightarrow I_\mathrm{late} = \frac{2i\omega_c}{m} \,  t_b + \cdots,
\end{equation}
where~$\cdots$ stand for contributions that are subleading in~$1/m$ and~$1/t_b$.  Working in the worldline approximation, this linear behavior at late time can be thought of as corresponding to a world line at fixed~ (generally complex) $w$ but extended in the~$t$ direction \cite{Hartman:2013qma}.  To identify these special values of~$w$ (which we denote~$w_c$), we extremize the action obtained from~\eqref{eq:action} by setting~$\dot{w}$ to zero. The resulting action is
\begin{equation}
\label{eq:Iaccumulate}
I_\mathrm{late} = \frac{i \ell}{z_0} \, \int V(w) \, dt,
\end{equation}
where
\begin{equation}
\label{eq:V}
V(w) = \sqrt{\frac{f(w)}{w}} - \Q(1-w),
\end{equation}
and its extremization requires solving~$V'(w) = 0$.   Finding the roots $w_c$ of~$V'(w)$ for general~$\alpha$,~$\qt$ amounts to solving the same sextic polynomial.   But since from~\eqref{eq:wdot} they satisfy~$g(w_c) = 0$, the corresponding energies are just the six $\E_c$ defined above:
\begin{equation}
\label{eq:Estarw}
\E_c = V(w_c).
\end{equation}
Thus in the late time limit we may write
\begin{equation}
\label{eq:omegaE}
I_\mathrm{late} = -\frac{2 i \ell \E_c}{z_0} t_b + \cdots \Rightarrow \omega_c = -m \left(E_E\right)_c,
\end{equation}
where $\left(E_E\right)_c = \ell \E_c/z_0$ is again the Euclidean energy from \eqref{consE}. Since the large time behavior of a physical probe field is controlled by its lowest QNM, we identify the frequency of this mode as $\omega_c$.  Thus, up to a factor of $m$, the critical energies~$\left(E_E\right)_c$ are directly related to such frequencies. In particular, for stable situations families of worldlines relevant at late times can have $t_b \rightarrow +\infty$ only for $\E_c$ in the upper half-plane.

Not only does the physics of QNMs determine the branch points $\E_c$, it also selects a physically meaningful location at which to place associated branch cuts~\cite{Festuccia:2005pi}.  This point may be seen by writing the  Fourier transform of the worldline-approximation correlator in the form
\bea
G_{12}(\omega) &\sim \int d\left(\Delta t\right) e^{-i\omega \Delta t} e^{-m I(\Delta t)}, \\
                &\sim \int dE_E \, e^{-m(I(E_E) + i(\omega/m)\Delta t(E_E))}.
\eea
For large~$m$, the dominant contribution to this integral comes from those~$E_E$ which satisfy the saddle point condition
\begin{equation}
\label{eq:saddle}
\frac{dI}{dE_E} + \frac{i\omega}{m} \, \frac{d(\Delta t)}{dE_E} = 0.
\end{equation}
Since $I$ is an action, we have the Hamilton-Jacobi relation
\begin{equation}
\label{eq:partialIt}
dI = iE_E \, d(\Delta t),
\end{equation}
which can also be checked directly from \eqref{eqs:DeltatI}.  Thus \eqref{eq:saddle} becomes simply~$E_E = -\omega/m$ at all times and the frequency space correlator is
\begin{equation}
\label{eq:OOfreq}
G_{12}(\omega) \sim e^{m Z(\omega)}, \hspace{0.5cm} \mbox{with } Z(\omega) = \left( i E_E\Delta t(E_E) - I(E_E)\right) \big|_{E_E = - \omega/m}.
\end{equation}
Since~$Z(\omega)$ and~$I(\Delta t)$ are related by a Legendre transformation, the analytic structure of the functions~$I$ and~$\Delta t$ in the complex energy plane is directly related to the analytic structure of the frequency space correlator $G_{12}(\omega)$.

In particular, the only singularities of the exact Green's functions $G_{12}(\omega)$ at finite $m$ computed using field theory should be poles corresponding to quasi-normal modes.  In the large $m$ limit these poles organize themselves into closely spaced families that define curves in the complex $\omega$ plane.  The endpoints of such curves are (some of) our $\E_c$'s and the associated lines of poles become branch cuts.    The most relevant observation is that the actual finite $m$ correlators are free of branch points, so that the parts of our Riemann surfaces $I(\E)$ and $\Delta t(\E)$ beyond the lines of poles are related much less directly to the physics of finite $m$.  In particular, even at general complex $t_b$ or $\omega$, finite $m$ correlators will never be well-approximated by $e^{-mI}$ for worldlines described by points behind such lines.

We henceforth restrict discussion to what we may call the principal sheets of $w_t(\E)$, $I(\E)$, and $\Delta t(\E)$ defined by introducing branch cuts along the large-$m$ lines of poles in $G_{12}(\omega)$.  We also take our principal sheets to include worldlines on which $\E$, $w_t$, are real and $\Delta t$ is purely imaginary. While determining the precise location of these cuts would require one to compute the full set of QNMs at large $m$, it will be enough for our purposes to note that QNMs typically become highly damped away from the lowest QNM.  Thus the branch cuts that determine our principal sheet must point away from the real~$\E$ axis.  A rough guess as to the appearance of these branch cuts is sketched in Figure~\ref{fig:Estar}. In particular, comparison with Figure~\ref{fig:chargedcurves} indicates that while all six values of $\E_c$ define branch points of the principal sheet for small~$\qt <\sqrt{2/3}$, at some point before $\qt = \sqrt{2/3}$ two of the $\E_c$ move onto a secondary sheet so that only the remaining four define branch points of the principal sheet and correspond to physical low-lying QNMs.  As one might expect, for $\qt > \sqrt{2/3}$ these are the four $\E_c$ with non-vanishing imaginary part.

\subsection{Correlators in the extreme limit}
\label{sec:full}

\begin{figure}
\centering
        \subfloat[\label{subfig:Eplanar}]{
        \includegraphics[page=11]{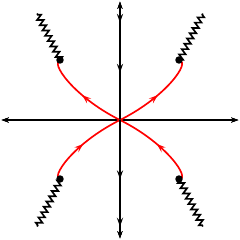}
        }
        \hspace{0.75cm}
        \subfloat[\label{subfig:Ealphasmall}]{
        \includegraphics[page=12]{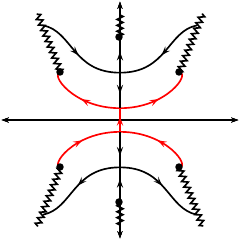}
        }
        \hspace{0.75cm}
        \subfloat[\label{subfig:Ealphabig}]{
        \includegraphics[page=13]{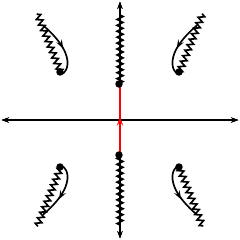}
        }
\caption{Sketches of the~$\E(t_b)$ contours corresponding to real values of~$t_b$ for a representative sample of~$\alpha$ with~$\qt = 0$; arrows on the contours indicate the direction of increasing~$t_b$.  Figure (a) is Schwarzschild, with $\alpha$ increasing to the right.  Taking $\alpha > 0$ introduces two additional~$\E_c$ along the imaginary axis and resolves the bifurcation point at~$\E = 0$. Note the presence of new contours that come in from infinity; following these contours to large $t_b$ takes us cross branch cuts (for which our rough guesses are shown as jagged lines) and off the principal sheet.  At $\alpha =\alpha_\mathrm{crit} \approx 0.406$ the bifurcation points merge and the contour topology changes to that of figure~(c).  We see that at least parts of some (black) contours move off the principal sheet. Red contours are associated with correlations that decay away from $t_b=0$ as in comment (ii).
}
\label{fig:neutralcurves}
\end{figure}

\begin{figure}[p]
\centering
\includegraphics[page=14]{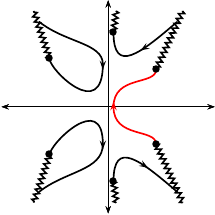}
\hspace{0.5cm}
\includegraphics[page=15]{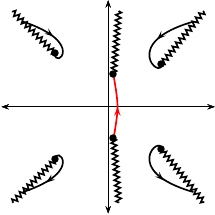}
\hspace{0.5cm}
\includegraphics[page=16]{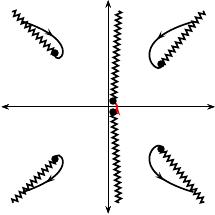}
\\
\vspace{0.5cm}
\includegraphics[page=17]{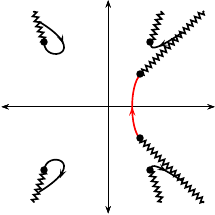}
\hspace{0.5cm}
\includegraphics[page=18]{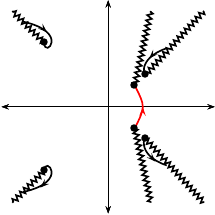}
\hspace{0.5cm}
\includegraphics[page=19]{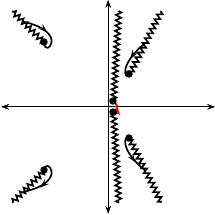}
\\
\vspace{0.5cm}
\includegraphics[page=20]{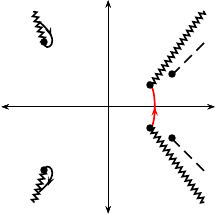}
\hspace{0.5cm}
\includegraphics[page=21]{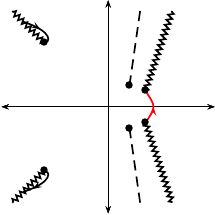}
\hspace{0.5cm}
\includegraphics[page=22]{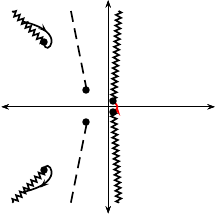}
\\
\vspace{0.5cm}
\includegraphics[page=23]{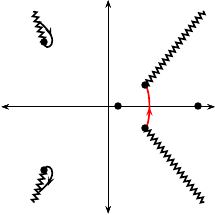}
\hspace{0.5cm}
\includegraphics[page=24]{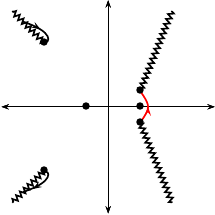}
\hspace{0.5cm}
\includegraphics[page=25]{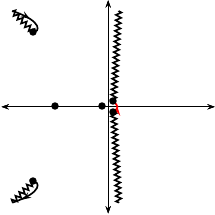}
\caption{Sketches of the~$\E(t_b)$ contours corresponding to real values of~$t_b$ for a representative sample of~$\alpha$ and~$\qt$; arrows on the contours indicate the direction of increasing~$t_b$. Red contours  are again associated with correlations that decay away from $t_b=0$ as in comment (ii).  The three columns take~$\alpha  = \alpha_1$,~$\alpha_2$,~$\alpha_3$, respectively, with~$\alpha_1 < \alpha_\mathrm{crit} < \alpha_2 < \alpha_3 \approx \alpha_\mathrm{ext}$; the rows from top to bottom take~$\qt = \qt_1$~$\qt_2$,~$\qt_3$,~$\qt_4$, respectively, with~$1 \gg \qt_1 < \qt_2 < \qt_3 < \sqrt{2/3} < \qt_4$.  As~$\qt$ is increased, two of the~$\E_c$ in the right-hand plane cross branch cuts (rough guesses for which are again shown as jagged lines), taking their associated contours and branch cuts off the principal branch; this is shown in the third row with such branch cuts indicated by dashed lines (suppressed in the 4th column).  At~$\qt = \sqrt{2/3}$, these two~$\E_c$ annihilate on the real axis; for~$\qt > \sqrt{2/3}$, these~$\E_c$ remain on the real axis and no contours on the principal sheet terminate on them.}
\label{fig:chargedcurves}
\end{figure}

As discussed for the Schwarzschild case in \cite{Fidkowski:2003nf}, it is generally quite subtle to determine which of the possible complex worldlines connecting our endpoints actually provides a good approximations to finite $m$ correlators via $G_{12} \sim e^{-mI}$.  The possible Schwarzschild contours $\E(t_b)$ for real $t_b$ are shown in Figure~\ref{subfig:Eplanar}.  While the correlator at~$t_b = 0$ corresponds to the unique~$\E = 0$ geodesic, this splits into three possible branches for nonzero~$t_b$.  By writing down a toy model for the path integral,~\cite{Fidkowski:2003nf} argued that the contours contributing to the path integral are the complex \linebreak \pagebreak \FloatBarrier \noindent ones terminating at~$\E_c = \sqrt{2} \, e^{i\pi/4 + k\pi/2}$ for~$k = 0,1,2,3$, while the contour along the  imaginary~$\E$ axis (which only reaches a finite value of~$t_b$ as~$\E \rightarrow i\infty$) does not contribute.  This is the case even though the imaginary $\E$ contour represents a smaller action for $t_b > 0$, and so would dominate if it contributed at all.  Adding charge ($\alpha > 0$) to the black hole and also to the probe ($\qt > 0$) leads to even more interesting structure for these contours which may further complicate the analysis.  See Figure \ref{fig:neutralcurves} for black holes with neutral probes and Figure \ref{fig:chargedcurves} for charged probes. The captions contain rough explanations of the evolution in $\alpha, \qt$, though due to our focus on the non-extreme case, we save further commentary for section~\ref{sec:nonextreme}.

Luckily, an indirect argument suffices to determine the correct contour in the extreme limit.  To see this,
recall from section \ref{sec:TFDCFT} that this limit must make our correlators independent of $t_b$.  Since $\Delta t = -2t_b +i \beta/2$, applying equation \eqref{eq:partialIt} to any contributing saddles requires $\E(t_b)$ to vanish at extremality for all $t_b$.  Taking the~$\eps \rightarrow 0$ limit of~\eqref{eq:Estarw} shows that that for~$\qt < 1$, precisely two critical energies $\E_c$ vanish at extremality; these are
\begin{equation}
\label{eq:Ec}
\E_c = \frac{\qt \pm i \sqrt{1-\qt^2}}{2\sqrt{3}} \, \eps + \mathcal{O}(\eps)^2 \hspace{0.5cm} (\qt < 1).
\end{equation}
One lies in the lower half plane, and the other lies in the upper half plane.  So for the stable case $\qt < 1$ there should be a unique contour connecting the two, and which should flow from the former to the latter. This is precisely what one finds numerically; see Figure \ref{fig:chargedcurves}. In the extreme limit, consistency thus requires that correlators receive contributions {\it only} from this contour.

Before analyzing this contour in detail, we remark that it displays several additional pleasing features:

\begin{enumerate}[i)]
\item{}  At least for $0 \le \qt \le 1$, for all nonextreme black holes with sufficient charge ($\alpha^2$ close enough to $2$) the $\E_c$ values corresponding to endpoints of the chosen contour continue to be the closest $\E_c$ to the real axis; see Figures \ref{fig:neutralcurves} and \ref{fig:chargedcurves}.  Thus, if they contribute at all, they give the lowest quasi-normal modes.

\item{}  It is natural to expect $|\qt| \le 1$ correlators to be largest at $t_b=0$ and to decay toward both future and past.  From \eqref{eq:partialIt}, this requires any dominant worldline at $t_b=0$ to have $\E$ real\footnote{Unless it somehow fails to contribute at all to any correlator with $t_b > 0$ (which would allow $\Im(\E) >0$), or to any correlator with $t_b < 0$ (which would allow $\Im(\E) >0$).  This seems unlikely even at special values of $\alpha, \qt$, and completely implausible on open sets of these parameters. The real $\E$ requirement applies also to cases where multiple worldlines share dominance at $t_b=0$.},   and also requires the contour in its vicinity to flow toward the upper half plane.

    This expectation is trivially satisfied for our chosen contour at $\alpha^2 =2$, $\qt =1$ for which $\E =0$ identically.  But the $t_b=0$ worldline on this chosen contour admits a unique continuous deformation to general allowed $\alpha, \qt$.  In each case one finds the corresponding $\E(t_b=0)$ to be real and the contour to flow in the desired direction.  This seen from Figures \ref{fig:neutralcurves} and \ref{fig:chargedcurves}, recalling that time-reversal symmetry relates the upper and lower half-planes. Thus if any contour crosses the real axis at a single point, this point must be $t_b=0$.  The relevant contour is shown in red for all $\alpha, \qt$.

\item{} The observation that following a branch cut from any $\E_c$ at fixed $\alpha, \qt$ should take one away from the real axis suggests near extremality that at least a large part of the other contours (for which $\E_c$ does not become small) do indeed move off the principal sheet.  See Figures \ref{fig:neutralcurves} and  \ref{fig:chargedcurves}.  It is plausible that at extremality such contours move off the principal sheet for all $t_b$, though this would require further analysis to determine.
\end{enumerate}

We also mention a further good property of our chosen contour for $\qt < 1$. Here the behavior near extremality is clear from general considerations even at the quantitative level. The RNAdS${}_5$ black hole develops a deep AdS${}_2 \times {\mathbb R}^3$ throat and, since this near-horizon region is associated with low energies, the emergent AdS${}_2$ isometries define an infrared conformal fixed point that associates each operator with a new effective infrared conformal dimension $\Delta_\mathrm{IR}$~\cite{Gubser:2008px,Gubser:2008pf}.  As discussed in \cite{Hartnoll:2008vx,Hartnoll:2008kx}, at finite but very low temperature to good approximation this is just a finite-temperature version of the same AdS${}_2$.  As a result, if we start with two points at the bifurcation surface and move them radially outward into opposite asymptotic regions then it is clear that $\Delta_\mathrm{IR}$ also controls the rate of decay of the associated two-point function.  Since this decay must continue up to a cutoff controlled by the temperature we arrive at
\begin{equation}
\label{eq:finiteDdecay}
G_{12} \sim T^{2\Delta_\mathrm{IR}}.
\end{equation}
In particular,  the precise condition for a non-vanishing two-point function as $T \rightarrow 0$ is $\Delta_\mathrm{IR} =0$, which indeed implies that the system sits on the threshold of an instability as discussed in \cite{Faulkner:2009wj}.

Returning to our contour, one can readily see that it provides results consistent with \eqref{eq:finiteDdecay}.  Since $\E \sim 0$ and $w_t \sim 1$, the leading-order behavior of~$I$ at fixed $t_b$ is
\begin{equation}
I = \ell \int^{1-\eps} \sqrt{\frac{1-\qt^2}{3}} \, \frac{1}{1-w} \, dw  + \mathcal{O}(1) = \ell \sqrt{\frac{1-\qt^2}{3}} \ln\left(\mu/T\right) + \mathcal{O}(1).
\end{equation}
Exponentiating this result gives \eqref{eq:finiteDdecay} with $\Delta_\mathrm{IR} = m \ell \sqrt{(1 - \qt^2)/12}$, which as one may easily check is the correct result at large $m$ in $5$ bulk dimensions.

We now consider $\qt =1$.  Since $\Delta_\mathrm{IR} =0$ to the approximation with which we work, the correlator will be finite at extremality.  But our worldline approximation provides interesting information about time-dependence.  To extract this information, we take the extreme limit~$\eps \rightarrow 0$ of~\eqref{eq:Estarw} in which three of the~$\E_c$ move to the origin.  Two of these correspond to the endpoints of our chosen~$\E \rightarrow 0$ contour. They are
\begin{equation}
\label{eq:Estarextremal}
\E_c = \left(\frac{1}{2\sqrt{3}} - \frac{e^{\pm i\pi/3}}{8 \cdot 3^{1/3}} \,  \eps^{1/3} \right) \eps + \mathcal{O}\left(\eps^{5/3}\right),
\end{equation}
suggesting that the entire~$\E(t_b)$ contour obeys the scaling relation
\begin{equation}
\label{eq:scaling}
\E = \left(\frac{1}{2\sqrt{3}} - a \, \eps^{1/3} \right) \eps
\end{equation}
with~$a$ a complex number ranging between~$e^{\pm i \pi/3}$.  Inserting this ansatz into the elliptic integral expressions of appendix \ref{app:elliptic} leads to simplified expressions (\eqref{eq:Deltata} and~\eqref{eq:Ia}) which satisfy the relation for all~$t_b$\footnote{To this order, the time dependent part of this result can also be obtained by using the leading ($a$-independent) term in~\eqref{eq:scaling} to integrate~\eqref{eq:partialIt}.}
\begin{equation}
\label{eq:IDeltat}
I = \ell\left[- \frac{\pi}{2\sqrt{3}} - 2\sqrt{\frac{2}{3}}\arctanh\sqrt{\frac{2}{3}} - \frac{2\pi i}{\sqrt{3}} \, T\, t_b + \mathcal{O}\left(T/\mu\right)^{4/3}\right],
\end{equation}
where the $\mathcal{O}(T/\mu)^{4/3}$ term has unspecified time dependence.  While this expansion can be continued to higher orders, it is simpler to focus on the late-time behavior and use~\eqref{eq:omegaE} and~\eqref{eq:Estarextremal}.  For example, we find
\begin{equation}
\label{eq:omegaextremal}
\omega_c = m\left[-\frac{\pi}{\sqrt{3}} \, \ell \, T + \frac{e^{-i\pi/3}}{2^{3/2}\cdot 3^{1/6}} \frac{\ell \mu}{\gamma} \left(\gamma \pi T/\mu\right)^{4/3} + \mathcal{O}(T/\mu)^{5/3}\right].
\end{equation}
The interesting property of both \eqref{eq:IDeltat} and \eqref{eq:omegaextremal} is that, beyond linear order in $T$, the expansion comes in powers of $(T/\mu)^{1/3}$.  This differs markedly from the $\qt < 1$ expansion which involves only integer powers of $T/\mu$.

\subsection{Comments on non-extreme contours}
\label{sec:nonextreme}

We now make some brief remarks on contours for general non-extreme $\alpha, \qt$ which, while tangential to our analysis of the extreme limit,  may nevertheless be of interest.

\begin{enumerate}[i)]
\item{}
The red contours in Figures \ref{fig:neutralcurves} and \ref{fig:chargedcurves} are associated with correlations that decay away from $t_b=0$ as in comment (ii).  We expect these to dominate for all $\alpha$ near $t_b =0$, though not necessarily for large $t_b$. Indeed, there is a regime between Figures \ref{subfig:Ealphasmall} and \ref{subfig:Ealphabig} where the red contour would reach the imaginary $\E_c$, while the complex $\E_c$ (off the imaginary axis) have smaller imaginary part, potentially corresponding to lower QNMs, and may well still contribute.

\item{}   The exchange of relevance/dominance of the above $\E_c$ near $\alpha_\mathrm{crit}$ in Figure \ref{fig:neutralcurves} appears to be related to the massless uncharged scalar results of \cite{Wang:2004bv} which found two families of QNMs: purely damped modes, and oscillating damped modes. The latter give the lowest QNM for small $\alpha$, while the former do so beyond some threshold value. Furthermore, the damping time diverges in the extreme limit. Though \cite{Wang:2004bv} studied global RNAdS$_4$, their results persist in the planar (large radius) limit.  Note that we find similar behavior for sufficiently small $\qt > 0$, though this transition disappears at larger $\qt$.

\item{} The location of the bifurcation points in Figure~\ref{fig:neutralcurves} corresponds to the location of the maxima and minima of~$t_b$ along the imaginary~$\E$ axis found by~\cite{Brecher:2004gn}, though they did not follow the complex branches.
\end{enumerate}

%================================================================
\section{Discussion}
\label{sec:discussion}
%================================================================

We have studied the behavior of thermofield double states with chemical potential $\mu$ in holographic contexts dual to the two-sided planar Reissner-Nordstr\"om AdS${}_5$ black hole.  One copy of the CFT is associated with each boundary, and we have focused on correlations and entanglement between the two.  The deep throat that arises in the extreme limit of RNAdS  immediately implies that corresponding two-point functions of neutral operators vanish as $T \rightarrow 0$ at fixed $\mu$.  For the same reasons, the thermo-mutual information between strips (or other finite-sized regions) of size $L$ in the two CFTs vanishes at small $T$ unless $L$ diverges; see section \ref{sec:mutualinfo}.  Such results might at first seem to suggest that all localized measures of entanglement vanish in this limit.

However, we have shown that other localized measures behave differently.  One example is the thermo-mutual information \eqref{s+CFT} between a width $L$ strip in one CFT and the full second CFT. As discussed in section \ref{TMI0}, this remains non-zero as $T \rightarrow 0$ so long as $L > L_{s+\mathrm{CFT}}|_{T/\mu=0} \approx 1.05\, \gamma/\mu$.   Another example is the two-point function of appropriately-tuned charged scalar operators.  In the limit of large conformal dimensions, the required tuning in bulk language is $m \kappa = qg$, which in field theory terms at large $\Delta$ becomes $\Delta = 2|q|$ for e.g.  $N=4$ SYM when the $U(1)$ charge corresponds to a subgroup of the SO(6) R-symmetry.   But as explained in section \ref{sec:full}, a more complete characterization of the requirement is that the effective IR conformal dimension of the operator should vanish, so that the system sits just on the threshold of an instability.

In particular, we saw explicitly in the wordline approximation that CFT two-point functions $G_{12}(x_1,x_2)$ with $\Delta_\mathrm{IR}=0$ remain non-zero at finite arguments and that the correlations they measure do not all shift off to infinitely large scales as $T \rightarrow 0$.  Since $\Delta_\mathrm{IR}$ controls the scaling of $G_{12}$ at any fixed spatial separation $\vec x_1 - \vec x_2$, we expect this behavior to continue even for finite-dimension operators; i.e., it is not an artifact of the worldline approximation.

We find the mixture of divergent and finite length scales as $T \rightarrow 0$ quite interesting.  The AdS${}_2 \times {\mathbb R}^3$ IR fixed point exhibits local criticality, with infinite dynamical scaling exponent $z$.  Since dynamical scaling symmetry at finite $z$ would require length scales $L \propto T^{-1/z}$, both constant $(T^0)$ and logarithmic behaviors ($\ln T$) are natural at $z =\infty$.  We see that AdS${}_2 \times {\mathbb R}^3$ fixed points involve a particular combination of the two  -- a result one would like to understand from the CFT perspective.  Motivated by our results for charged correlators, one would also like to study what one might call the ``charged thermo-mutual information'' of two finite strips (one in each CFT) defined using the von Neumann version of the charged R\'enyi entropies of \cite{Belin:2013uta}.  At least with additional fine-tuning of the new charge parameter, this may well lead to further entanglement measures that remain localized as $T \rightarrow 0$.

Understanding the entanglement structure of physically interesting states at various scales is an intriguing and complex problem.  Indeed, this is the goal of many studies of tensor network representations of ground states, the multi-scale entanglement renormalization ansatz (MERA), and the like; see e.g. \cite{Ostlund1995,Verstaete2004,Vidal2008,Vidal2009,schollwock2011}.  Our parameter $T/\mu$ is a dial that one can turn to explore this scale-dependence for TFD states at $t=0$, just as one may explore the time-dependence of entanglement using the proposal of \cite{Hubeny:2007xt} (see e.g. \cite{Hubeny:2007xt,Calabrese:2009qy,AbajoArrastia:2010yt,Albash:2010mv,Balasubramanian:2010ce,Asplund:2011cq,Aparicio:2011zy,Basu:2011ft,Balasubramanian:2011at,Allais:2011ys,Basu:2012gg,Buchel:2013lla,Hartman:2013qma,Liu:2013iza,Liu:2013qca}).  The two limits are closely related, as both explore the deep infrared.  Indeed, our results for TMI at low $T$ have much in common with those of \cite{Hartman:2013qma} at late times: There again the thermo-mutual information vanished between strips of fixed finite size in opposing CFTs, while  -- although not actually discussed in \cite{Hartman:2013qma} -- TMI(${\cal A}_1$:CFT${}_2$) need not vanish since it is in fact independent of time.

As noted earlier, such observations are difficult to reconcile with the quasiparticle picture of TMI entanglement (see \cite{Hartman:2013qma}, following the time-dependent picture of \cite{Calabrese:2009qy}).  In particular,
as described in section \ref{sec:s+CFT}, we find that the TMI between a strip in one CFT and its complement in the other again vanishes at small $T$, so that one cannot even say that the CFT${}_2$ degrees of freedom entangled with a given strip in CFT${}_1$ have moved off to infinite scales -- they remain tied in some essential way to the mirror-strip in CFT${}_2$.  The situation is even more dramatic if we compactify space, in which case
the analogues of ${\cal A}_1$, ${\cal A}_2$, ${\cal A}_1^c$, ${\cal A}_2^2$, all have pairwise vanishing TMI at sufficiently small $T$. And it is clear that this same behavior will be found at late times using \cite{Hubeny:2007xt}.   Assuming this prescription to be correct thus leads to a similarly dramatic late-time failure of the quasiparticle picture for any initial state\footnote{Other features of the proposal \cite{Hubeny:2007xt} that are diffcult to reconcile with a free-streaming quasiparticle picture of time-dependence were mentioned in \cite{Liu:2013iza,Liu:2013qca}.  We comment that CFTs on spaces with compact directions provide yet another.  For example, in a $d=2$ CFT on a circle of radius $R$, all quantities associated with free-streaming quasiparticles of speed $v$ are periodic with period $2\pi R/v$.  But aside from trivial conserved currents at the boundary, holographic duals certainly do not display this periodicity, and neither should more general CFTs.}.

A final general feature on which we remark is the sharpness of transitions associated with TMI at large $N$, in that it strictly vanishes below some threshold.  Such sharpness is of course a general feature of transitions involving holographic entanglement \cite{Ryu:2006bv}.  The fact that this behavior is by now well-known should not reduce our desire to understand it at the microscopic level.  Indeed, it seems deeply related to the general observation that plasmas in holographic CFTs can strongly \textit{de}couple from short-distance probes.  A particularly striking example of such behavior is the funnel/droplet transition described in \cite{Hubeny:2009ru} -- see \cite{Marolf:2013ioa}  for a review -- in which such plasmas suddenly become unable to couple to heat sources smaller than some characteristic size.   The funnel/droplet transition was recently linked to color confinement \cite{Marolf:2013ioa}, and the resulting circle of ideas may have implications for the present discussion.

Let us now briefly return to two-point functions.  In addition to the results summarized above, we also found new phenomena associated with ``extremally charged'' operators ($\qt =1$) at small $T$.   In particular, since all time-dependence of the $T=0$ TFD is through an overall phase, physical quantities become time-independent. But at least for operators tuned to satisfy $\Delta_\mathrm{IR}=0$, the precise way in which they do so seems to be via an unexpected expansion in powers of $T^{1/3}$ that governs corrections beyond the leading linear behavior (see equation \eqref{eq:omegaextremal}).  While section \ref{sec:mutualinfo} reported TMI results only for $t=0$, using the proposal of \cite{Hubeny:2007xt} the analysis extends readily to more general times and produces late-time results that agree with \cite{Liu:2013iza,Liu:2013qca} and which give only smooth functions of $T$.

As a final comment, we recall that~\cite{Fidkowski:2003nf} described how two-point TFD correlators similar to those studied here might be used to probe the classical singularity of the planar Schwarzschild solution ($\mu =0$), and thus perhaps to study how this singularity is resolved by quantum and/or stringy effects.  While already nontrivial at $\mu =0$, we note that any generalization to $\mu \neq 0$ will involve further subtleties.  In particular, for $\mu =0$ the idea was to study operators of large but finite dimension and to analytically continue $t_b$ until the associated geodesic passes close to the singularity -- in our notation, until $w$ becomes very large.  As is clear from the upper left diagram in Figure~\ref{fig:wbranches}, for $\mu =0$ this happens as $\E \rightarrow \pm i \infty$ along the principal sheet of the $w_t$ Riemann surface.  But as shown in the other diagrams in Figure~\ref{fig:wbranches}, for $\mu \neq 0$ one finds that $w_t$ remains bounded on the principal sheet. Thus finite $m$ correlators are no longer approximated by geodesics passing close to the singularity anywhere in the complex $t_b$ plane.  The construction analogous to~\cite{Fidkowski:2003nf} would thus require first taking the $m \rightarrow \infty$ limit of finite-dimension correlators and then analytically continuing to another sheet of the Riemann surface $w_t(\E)$.  Indeed, from our preliminary numerics it is unclear whether one can even reach the inner horizon on the principal sheet, so this same complication may well apply to analogous investigations of inner horizon instabilities along the lines of \cite{Brecher:2004gn}.

%================================================================
\section*{Acknowledgements}

We thank David Tong, Tom Hartman, Veronika Hubeny, Mukund Rangamani, Harvey Reall, Jorge Santos, Toby Wiseman, Jerome Gauntlett, Kostas Skenderis, and Benson Way for useful discussions.  TA and SFR were supported by the STFC. SF and DM  were supported in part by the National Science Foundation under Grant No PHY11-25915, by funds from the University of California.  DM also received support as a Visiting Fellow College at Trinity College, University of Cambridge, UK.  MR is supported by a Discovery Grant from NSERC\ of Canada. SF and DM also thank the Department of Applied Mathematics and Theoretical Physics for their hospitality during the bulk of this work. Finally, we also gratefully acknowledge both the Centro de Ciencias de Benasque Pedro Pascual and the Isaac Newton Institute for their hospitality during various stages of this work.

\appendix

\section{Geodesic Approximation for Charged Operators}
\label{app:action}

Here we derive the form of the action~\eqref{eq:action} in the limit of large~$m$, following~\cite{Louko:2000tp}.  The Green's function for the field~$\phi$ with mass and charge~$m$ and~$q$ should be a Green's function of the Klein-Gordon operator~$H = (-i\partial - qA)^2 + m^2 = (p-qA)^2 + m^2$; we can represent this Green's function as
\begin{equation}
\frac{-i}{H} = \int_0^\infty e^{-iNH} \, dN,
\end{equation}
so that using the standard path integral construction, we get
\begin{equation}
\left\langle x \middle| \frac{-i}{H} \middle| y \right\rangle = \int_0^\infty dN \, \int \mathcal{D}x \, \mathcal{D}p \, \exp\left\{i\int_0^1 \left[\dot{x} p - N((p - qA)^2 + m^2) \right] d\lambda \right\}.
\end{equation}
We can interpret~$N$ as a field in some appropriate gauge-fixing; we can make this explicit by introducing the gauge-fixing condition and determinant.  Then we obtain
\begin{equation}
\left\langle x \middle| \frac{-i}{H} \middle| y \right\rangle = \int \mathcal{D}N \, \mathcal{D}x \, \mathcal{D}p \, \Delta(x) \, \exp\left\{i\int_0^1 \left[\dot{x} p - N((p - qA)^2 + m^2) \right] d\lambda \right\},
\end{equation}
where now~$N$ is a field to be integrated over.  Now, in the WKB approximation, we can integrate out the fields~$N$ and~$p$ by replacing them in the action with their on-shell values.  Their equations of motion are
\begin{subequations}
\bea
(p-qA)^2 + m^2 &= 0, \\
\dot{x} - 2N(p-qA) &= 0,
\eea
\end{subequations}
so their on-shell values are
\begin{subequations}
\label{eqs:pN}
\bea
p &= \frac{m\dot{x}}{\sqrt{-\dot{x}^2}} + qA, \\
N &= \frac{\sqrt{-\dot{x}^2}}{2m}.
\eea
\end{subequations}
The correlator then becomes
\begin{equation}
\left\langle x \middle| \frac{-i}{H} \middle| y \right\rangle = \int \mathcal{D}x \, (\cdots) \exp\left\{-\int_0^1 \left[m\sqrt{\dot{x}^2} - iqA \dot{x} \right] d\lambda \right\},
\end{equation}
where~$(\cdots)$ represents functional determinants that we can neglect at leading order in the WKB approximation.  Approximating the path integral over~$x$ using the saddle point method, we get
\begin{equation}
\left\langle x \middle| \frac{-i}{H} \middle| y \right\rangle \sim e^{-mI[x_\mathrm{cl}]},
\end{equation}
where~$x_\mathrm{cl}$ is a solution to the equations of motion that come from the action
\begin{equation}
I[x] = \int \left[\sqrt{\dot{x}^2} - \frac{iq}{m} A \dot{x} \right] d\lambda.
\end{equation}
This last expression is precisely~\eqref{eq:action} used in the text.  Note that it differs from the action used in \cite{Brecher:2004gn} by a crucial factor of $i$ in the second term.

\section{Evaluation of the Elliptic Integrals}
\label{app:elliptic}

Our notation in this appendix follows \cite{elliptic}.  The expressions for~$\Delta t$ and~$I$ in terms of elliptic integrals are
\begin{subequations}
\begin{multline}
\Delta t = \frac{2 z_0 \Q}{\alpha^2 \sqrt{\alpha^2 - \Q^2}} \frac{1}{\sqrt{\Delta_{21}}} \left\{\frac{\Delta_{20}}{(w_2-1)\Delta_{2-}\Delta_{2+}} \left[F(\psi|m)-K(m)\right] \right. \\
 \left. + \frac{1-w_0}{(1-w_-)(w_+-1)(w_2-1)} \left[\Pi\left(\frac{w_2-1}{\Delta_{21}};\psi \middle| m\right) - \Pi\left(\frac{w_2-1}{\Delta_{21}} \middle| m\right)\right] \right. \\
  \left. + \frac{\Delta_{0-}}{(1-w_-)\Delta_{+-}\Delta_{2-}} \left[\Pi\left(\frac{\Delta_{2-}}{\Delta_{21}};\psi \middle| m\right) - \Pi\left(\frac{\Delta_{2-}}{\Delta_{21}} \middle| m\right)\right] \right. \\
\left. + \frac{\Delta_{0+}}{(w_+-1)\Delta_{+-}\Delta_{2+}} \left[\Pi\left(\frac{\Delta_{2+}}{\Delta_{21}};\psi \middle| m\right) - \Pi\left(\frac{\Delta_{2+}}{\Delta_{21}} \middle| m\right)\right]\right\},
\end{multline}
\begin{multline}
I = \frac{2i}{\sqrt{\Delta_{21}}} \left\{-\frac{h(w_2)}{w_2\Delta_{2-}\Delta_{2+}} \left[K(m) - F(\psi|m)\right] \right. \\
\left. - \frac{h(w_-)}{w_-\Delta_{2-}\Delta_{+-}}\left[\Pi\left(\frac{\Delta_{2-}}{\Delta_{21}}\middle| m \right) - \Pi\left(\frac{\Delta_{2-}}{\Delta_{21}};\psi \middle| m \right)\right]
\right. \\
\left. + \frac{h(w_+)}{w_+\Delta_{2+}\Delta_{+-}}\left[\Pi\left(\frac{\Delta_{2+}}{\Delta_{21}}\middle| m \right) - \Pi\left(\frac{\Delta_{2+}}{\Delta_{21}};\psi \middle| m \right)\right] + \frac{h(0)}{w_2 w_+ w_-} \Pi\left(\frac{w_2}{\Delta_{21}} \middle| m\right)\right\} \\
+ \frac{2ih(0)}{\sqrt{\Delta_{21}} \, w_2 w_+ w_- } \Pi\left(\frac{w_2}{\Delta_{21}}; \arctan\sqrt{\frac{\Delta_{21}}{w_1-w_\mathrm{UV}}} \middle| m\right)+ I_\mathrm{ct},
\end{multline}
\end{subequations}
where
\begin{subequations}
\bea
\tan\psi &\equiv \sqrt{\frac{\Delta_{21}}{w_1}}, \\
m &\equiv \frac{\Delta_{23}}{\Delta_{21}}, \\
\Delta_{ij} &\equiv w_i - w_j, \\
w_0 &\equiv 1+\frac{\E}{\Q}, \\
h(w) &\equiv \frac{\ell}{\alpha^2\sqrt{\alpha^2-\Q^2}}\, \left[1+w-\alpha^2 w^2 - \Q w(\E + \Q(1-w))\right].
\eea
\end{subequations}
These expressions have branch points wherever~$m = 1$ or~$\infty$, corresponding to points where~$w_1 = w_2$ or~$w_1 = w_3$.

In the scaling limit~\eqref{eq:scaling} with~$\qt = 1$, the above expressions reduce to
\begin{equation}
\label{eq:Deltata}
\frac{2\Delta t}{\beta} = -\frac{8\cdot 3^{1/4} \, i\, b^{2/3}\left[K(\widetilde{m}) - \Pi(n|\widetilde{m})\right]}{\pi\left(8\cdot 3^{2/3} e^{i\pi/3} a - \sqrt{3} \, e^{-i\pi/3} b^{2/3}\right)\sqrt{16\cdot 3^{1/6} e^{-5i\pi/6} \, a \, b^{1/3} + 2e^{-i\pi/6}b}},
\end{equation}
\begin{multline}
\label{eq:Ia}
I = \ell \left\{\frac{4 \, b^{2/3}\left[K(\widetilde{m}) - \Pi(n|\widetilde{m})\right]}{3^{1/4}\left(8\cdot 3^{2/3} e^{i\pi/3} a - \sqrt{3} \, e^{-i\pi/3} b^{2/3}\right)\sqrt{16\cdot 3^{1/6} e^{-5i\pi/6} \, a \, b^{1/3} + 2e^{-i\pi/6}b}} \right. \\
\left. - 2\sqrt{\frac{2}{3}}\arctanh\sqrt{\frac{2}{3}}\right\},
\end{multline}
where
\begin{subequations}
\label{eqs:params}
\bea
b &= 1+\sqrt{1+512\sqrt{3} \, a^3}, \\
\widetilde{m} &= \frac{\sqrt{3}(8 \cdot 3^{1/6} a + b^{2/3})}{8\cdot 3^{2/3} e^{-i\pi/3} a + \sqrt{3} \, e^{i\pi/3}b^{2/3}}, \\
n &= \frac{8\cdot 3^{1/6} e^{-i\pi/6} a + e^{i\pi/6} b^{2/3}}{8\cdot 3^{2/3} e^{-i\pi/3} a + \sqrt{3} \, e^{i\pi/3}b^{2/3}}.
\eea
\end{subequations}

Next, consider the indefinite version of the integral \eqref{A conn 1},
\begin{equation}\label{ellip int cal I}
        {\cal I}_\alpha(w)  = \frac{i}{2 \alpha} \int \frac{dw}{w^2 [ (1-w) (w - w_+) (w - w_-)  ]^{1/2}}.
\end{equation}
We are interested in the logarithmic divergence that comes from $w = 1$ for small $\epsilon = 2 - \alpha^2$.
We can extract it by noting that \eqref{ellip int cal I} can be written in terms of Elliptic integrals as
\begin{align}
\nonumber
        \frac{2 \alpha}{i}{\cal I}_\alpha(w) &= - \frac{1}{w w_-} \left( \frac{(1 - w) (w-w_-) }{ w-w_+} \right)^{1/2}
        - \frac{ (w_+-1)^{1/2}  }{w_+ w_-} E( \hat \psi | \hat m )  \\
\label{cal I 2}
        &+ \frac{(w_+ + 1)}{w_+^2 (w_+-1)^{1/2}} F( \hat \psi |  \hat m ) + \frac{[ w_- + w_+ ( w_- + 1)]}{ w_- w_+^2 (w_+ -1)^{1/2}} \Pi ( \hat n;  \hat\psi |  \hat m ),
\end{align}
\noindent where
\begin{equation}
        \tan \hat \psi =  \left( \frac{w_+ - 1}{1-w}  \right)^{1/2}, \qquad \hat m = \frac{w_+ - w_-}{w_+ -1}, \qquad  \hat n = \frac{w_+}{w_+ -1}.
\end{equation}
For small~$\eps$, the dominant contribution to the integral comes from $w = 1$, so we can drop the first term in \eqref{cal I 2} and let $\hat \psi = \pi/2$ for all $w_+ > 1$. Then the incomplete
Elliptic integrals reduce to complete ones, i.e. $F(\pi/2| x) = K(x)$, $E(\pi/2| x) = E(x)$, $\Pi(x; \pi/2 | y) = \Pi(x | y)$.
Using the asymptotics for small $z$
\begin{align}
        E\left ( \frac{a}{z} \right)  &= i \frac{a^{1/2}}{z^{1/2}} + O( z^{1/2}), \\
        K\left ( \frac{a}{z} \right)  &= - i \frac{z^{1/2}}{2 a^{1/2}} \log \left( - \frac{16 a}{z} \right) + O( z^{3/2}), \\
        \Pi \left (\frac{b}{z} \bigg | \frac{1}{z} \right ) &= \frac{z^{1/2}}{2(b-1)^{1/2}} \left[ \log \left(  \frac{\sqrt{b-1} + i}{\sqrt{b-1} - i} \right) - i \pi \right],
\end{align}
\noindent we arrive at \eqref{A conn app 2}.

%================================================================
% BIBLIOGRAPHY
%================================================================

\bibliographystyle{JHEP}
\bibliography{biblio}

\end{document}